\documentclass[usegraphicx,usenatbib,useAMS]{mn2e}
\usepackage{aas_macros}

\usepackage[pdfborder={0 0 0.1}]{hyperref} 

\newcommand\ion[2]{#1{\thinspace\scshape#2}}%

\newcommand\CIV{\ion{C}{iv}}
\newcommand\OmegaCIV{\ensuremath{\Omega_\mathrm{C\thinspace IV}}}
\newcommand\OmegaCII{\ensuremath{\Omega_\mathrm{C\thinspace II}}}
\newcommand\OmegaC{\ensuremath{\Omega_\mathrm{C}}}
\newcommand\OmegaCIVDot{\ensuremath{\dot{\Omega}_\mathrm{C\thinspace IV}}}
\newcommand\OmegaCIVInstant{\ensuremath{\Omega_\mathrm{C\thinspace IV inst}}}
\newcommand\OmegaCIVDelay{\ensuremath{\Omega_\mathrm{C\thinspace IV delay}}}
\newcommand\OmegaCIVDelayDot{\ensuremath{\dot{\Omega}_\mathrm{C\thinspace IV delay}}}
\newcommand\OmegaCIVInstantDot{\ensuremath{\dot{\Omega}_\mathrm{C\thinspace IV inst}}}

\title[Delayed Enrichment by Unseen Galaxies]{Delayed Enrichment by
  Unseen Galaxies: Explaining the Rapid Rise in IGM CIV Absorption
  from $z=6$--$5$}

\author[Kramer, Haiman \& Madau]{%
  R. H. Kramer \thanks{Zwicky Fellow, ETH Zurich, roban.kramer@phys.ethz.ch}$^{1}$,
  Z. Haiman \thanks{zoltan@astro.columbia.edu}$^{2}$,
  and P. Madau \thanks{pmadau@ucolick.org}$^{3}$
  \\
  $^{1}$ Institute for Astronomy, ETH Zurich, Wolfgang-Pauli-Strasse
  27, CH-8093 Zurich, Switzerland\\
  $^{2}$Department of Astronomy, Columbia University, 550 West 120th
  Street, New York, NY 10027, USA\\
  $^3$Department of Astronomy and Astrophysics, University of
  California Santa Cruz, 1156 High Street Santa Cruz, CA 95064, USA}

\begin{document}

\date{}

\pubyear{2010}

\maketitle

\begin{abstract}
In the near future, measurements of metal absorption features in the
intergalactic medium (IGM) will become an important constraint on
models of the formation and evolution of the earliest galaxies, the
properties of the first stars, and the reionization and enrichment of
the IGM. The first measurement of a metal abundance in the IGM at a
redshift approaching the epoch of reionization already offers
intriguing hints. Between $z=5.8$ and $4.7$ (a $0.3$~Gyr interval only
$1$~Gyr after the big bang), the measured density of \ion{C}{iv}
absorbers in the IGM increased by a factor of $\sim 3.5$
(\citealp{2009MNRAS.395.1476R}; \citealp*{2009ApJ...698.1010B}). If
these values prove to be accurate, they pose two puzzles: (1) The
total amount of \ion{C}{iv} at $z=5.8$ implies too little star
formation to reionize the IGM by $z=6$ or to match the {\it WMAP}
electron scattering optical depth ($\tau$). (2) The rapid growth from
$z \approx 6$--$5$ is faster than the buildup of stellar mass or the
increase in the star formation rate density over the same interval. We
show that a delay of $\sim 0.4$--$0.7$~Gyr between the instantaneous
production of ionizing photons and the later production of metal
absorption features (added to the delay due to stellar lifetimes) can
provide the full explanation for both puzzles. We calculate the delay
in metal production due to finite stellar lifetimes alone and find
that it is too short to explain the rapid \ion{C}{iv} density
increase. The additional delay could naturally be explained as the
result of $\sim 200~\mathrm{km}/\mathrm{s}$ outflows carrying carbon
to distances of $\sim 100$~kpc, the typical distance between galaxies
and \ion{C}{iv} absorbers in enrichment simulations, and the typical
outflow or absorption region scale observed at $z \approx 2$--$3$.
\end{abstract}


\section{Introduction}\label{sec_introduction}

Most of the star formation taking place during the reionization epoch
remains invisible to current observations, and a large fraction may
remain invisible even to the {\it James Webb Space Telescope (JWST)}
\citep*[e.g.][]{arXiv:1003.3873}. This unobserved population of faint
galaxies must have contributed most of the high--energy photons that
reionized the intergalactic medium (IGM). The ionization state of the
intergalactic medium is one important probe of this population. Other
constraints are provided by the detection of luminous Lyman--break
galaxies (LBGs) and Lyman--$\alpha$ emitters (LAEs) at $z>6$
\citep[e.g.][]{arXiv:1006.4360, arXiv:1006.3071}. Although the
directly observed population cannot account for all of the photons
necessary to reionize the IGM (and especially to match the {\it WMAP}
Thompson--scattering option depth from free electrons $\tau$,
\citealp{2007MNRAS.382..325B, 2009ApJ...690.1350O}), they constrain
the high--luminosity end of the galaxy luminosity function (LF).

The distribution (by element, in time, and in space) of the elements
synthesised in these early generations of stars and subsequently
expelled into the IGM (or incorporated into low--mass stars still
observable in the local universe) will become an important source of
information about the epoch of reionization in the near
future. Currently the highest--redshift measurements of metal
abundance in the IGM come from searches for \ion{C}{iv} absorption
features in quasar spectra (\citealp{2009MNRAS.395.1476R};
\citealp*{2009ApJ...698.1010B}).

The uncertainties on the $z \ga 5$ \ion{C}{iv} measurements are still
large, but if the values prove to be accurate then they pose two
puzzles. First, \citet{2009MNRAS.395.1476R} find that the total amount
of \ion{C}{iv} at $z=5.8$ implies too little star formation to
reionize the IGM by $z=6$. Second, \citet{2009ApJ...698.1010B} find
that the fractional increase in \ion{C}{iv} density is larger than
either the buildup of stellar mass or the increase in the star
formation rate density (SFRD) over the same interval.

We show here that both puzzles can be solved by a delay between the
production of ionizing photons and the enrichment of the IGM with the
associated metals. In the next section (\S \ref{sec_litreview}), we
discuss current observations and models of \ion{C}{iv} in the IGM in
more detail. Section \ref{sec_methods} presents our simple framework
for modelling reionization and enrichment based on a galaxy luminosity
function history.%
\footnote{The full compendium of code and ancillary files needed to
  reproduce the present paper is available from the first
  author. Cosmologial calculations were made with the
  \textsc{CosmoloPy} package. The \textsc{EnrichPy} package
  encapsulates our enrichment model.
  These resources are available at 
  \url{http://www.astro.phys.ethz.ch/kramer/},\\%
  \url{http://roban.github.com/CosmoloPy/}, and \\%
  \url{http://roban.github.com/EnrichPy/}.}
In Section \ref{sec_results}, we show that we can match reionization
(\S \ref{sec_reionresults}) and enrichment constraints at a single
redshift (\S \ref{sec_CIVresults}) with simple models and physically
plausible parameters, then go on to show that a delay of
$0.4$--$0.7$~Gyr (in addition to the delay due to finite stellar
lifetimes) can produce the observed rapid rise in \ion{C}{iv} absorber
density (\S \ref{sec_riseresults}). In Section \ref{sec_discussion},
we propose two explanations for the delay involving galactic outflows,
and suggest observational tests of those explanations. In Section
\ref{sec_conclusions}, we summarise our results and discuss how future
observations and models will improve our understanding of early metal
enrichment and reionization.

\section{Existing Observations and Models}\label{sec_litreview}

Observations of absorption features in high--redshift quasar spectra
are beginning to probe the IGM metallicity at redshifts approaching
$z=6$ (arguably close to the final stages of reionization). Currently
the highest--redshift measurements are of the \ion{C}{iv}
$1548.2$,~$1550.8$~\AA\ doublet, redshifted into the near infrared
(NIR). \citet{2009MNRAS.395.1476R} identified three (plus one
tentative) \ion{C}{iv} features in a search between $z=5.2$ and $6.2$
along lines of sight to $9$ quasars with a combined absorption
distance of $\Delta X = 25.1$.%
\footnote{$\Delta X$ is defined so that objects with constant comoving
  density and physical cross section have constant density per unit
  $\Delta X$ \citep{1988ApJ...329L..57T, 2009MNRAS.395.1476R}.}
\citet{2009ApJ...698.1010B} performed a similar search in four sight
lines ($\Delta X = 11.3$), finding no absorption features. This is
particularly surprising since their observations were sensitive to
even lower column densities than \citet{2009MNRAS.395.1476R}, and
therefore would have been expected to detect more \ion{C}{iv} features
per unit absorption distance if the column--density distribution
followed the declining power--law form found at lower redshift.

Converting the absorption feature detections to the average
\ion{C}{iv} density in the IGM (expressed as a fraction of the
critical density), and correcting for their completeness limits in
column density, \citet{2009MNRAS.395.1476R} found
\begin{equation}
 \OmegaCIV \left(\langle z \rangle = 5.76\right) = (5.0 \pm 3.0)\times
 10^{-9} \textrm{.}
\end{equation}
Applying similar corrections to observations by
\citet{2003ApJ...594..695P}, they found
\begin{equation}
 \OmegaCIV \left(\langle z \rangle = 4.69\right) = (17 \pm 6) \times
 10^{-9} \textrm{.}
\end{equation} 
The errors are still large on these measurements. If the $z=5.76$
value is revised toward the upper end of the allowed range, then the
puzzle of the rapid evolution in \ion{C}{iv} will be greatly reduced
(see Figure \ref{fig_deltas} in \S \ref{sec_results}). However we will
proceed here under the assumption that the central values are
essentially correct. We therefore seek to resolve the puzzle of the
rapid \ion{C}{iv} growth.

The \ion{C}{iv} density in the IGM is approximately constant at
$\OmegaCIV \sim 2 \times 10^{-8}$ from $z=2$--$4$
\citep{2009MNRAS.395.1476R,2001ApJ...561L.153S}, a surprising result
given that this is a period of intense star formation. Models of IGM
enrichment have successfully explained this as the result of a
decreasing fraction of carbon in the triply--ionized state, offsetting
the concurrent rise in total carbon density
\citep{2006MNRAS.373.1265O, 2007MNRAS.374..427D, 2008MNRAS.387..577O}.

Models of enrichment and carbon ionization at $z>4$ have been able to
produce a rise in \OmegaCIV{} between $z=6$ and $5$ consistent with
the observations \citep*{2009MNRAS.396..729O,arXiv:1005.1451}, at
least within their large uncertainties. Both
\citet{2009MNRAS.396..729O} and \citet{arXiv:1005.1451} find that the
total amount of carbon in the IGM increases by a factor of $2.5$ to
$3$ in this interval, while the fraction of carbon in the
triply--ionized state only increases by a factor of $\la 1.25$.

These simulations do not self--consistently model reionization, so
they are unable to directly elucidate the relationship between
reionization and enrichment, and therefore unable to provide a
satisfying solution to the first puzzle posed above, which is our
primary concern here. We are therefore interested in investigating the
connections between reionization, enrichment, and the population of
galaxies responsible for both.

\section{Methods}\label{sec_methods}

\subsection{Connecting Reionization and the Galaxy Luminosity Function}\label{sec_LFion}

In order to solve the two puzzles posed by the rapid \OmegaCIV{} rise
between $z=6$ and $5$ (\S \ref{sec_introduction}), we need to model
the luminosity function of galaxies, and from that model calculate the
total ionizing emissivity as a function of redshift. We assume the LF
is a Schechter function. Observations suggest that $\phi_*$ (the
density normalisation) and $\alpha$ (the faint--end slope of the LF)
are approximately constant from $z=4$ to $6$
\citep{2007ApJ...670..928B}. Neither are constrained as precisely at
higher redshift, but \citet{arXiv:1006.4360} find that the observed LF
at $z=7$ and $8$ is consistent with constant $\alpha$ and $\phi_*$,
though the maximum--likelihood values move toward steeper faint--end
slopes ($\alpha=-1.94\pm0.24$ at $z\sim7$ and $-2.00 \pm 0.33$ at
$z\sim8$). We therefore parameterise the evolution of the LF by
varying only $M_*$ with redshift. At $z < 9.0$ we use the $M_*$ values
from \citet{2008ApJ...686..230B}, linearly interpolated as a function
of redshift. At $z>9$ we assume $M_*$ is linear in redshift with slope
$\beta_{M*}$. The parameters $\alpha$ and $\beta_{M*}$ therefore
control the extrapolation of the LF to lower luminosities and higher
redshifts than have yet been probed by observations.

To convert the observed LF to an ionizing photon emissivity (or rate
density, $\textrm{photons}^{-1}~\textrm{s}^{-1}~\textrm{Mpc}^3$), we
use the \citet{2007MNRAS.382..325B} spectral energy distribution (SED)
and an escape fraction $f_{\mathrm{esc}\gamma}$. With this SED, the
ratio of ionizing photon production rate to UV luminosity is
\begin{equation}
\frac{\dot{N}}{L(1500~\textrm{\AA})} = 8.4\times 10^{24}~%
\frac{\textrm{photons}~\textrm{s}^{-1}}%
     {\textrm{erg}~\textrm{s}^{-1}~\textrm{Hz}^{-1}},
\end{equation}
though we assume only a fraction $f_{\mathrm{esc}\gamma}$ of these
escape into the IGM.%
\footnote{Note that $\dot{N}/L(1500~\textrm{\AA})$ is sensitive to the
  initial mass function and metallicity of the stellar population,
  though we ignore this dependence
  here. $\dot{N}/L(1500~\textrm{\AA})$ is completely degenerate with
  $f_{\mathrm{esc}\gamma}$ in our formalism.}

With these factors we convert the integrated LF history into an
ionizing emissivity history, and then into the ionized fraction of the
IGM $x(z)$.  We ignore twice--ionized helium, assume that
once--ionized helium has the same number fraction as hydrogen, and
include recombinations of hydrogen. Recombinations are calculated with
clumping factor $C \equiv \langle n_\mathrm{HII}^2 \rangle / \langle
n_\mathrm{HII} \rangle ^2 = 4$ (see \S \ref{sec_reionresults} for
discussion of varying $C$), gas temperature $10^4$~K (giving case B
recombination rate $\alpha_B = 2.6\times
10^{-13}~\mathrm{cm}^3~\mathrm{s}^{-1}$,
\citealp{1997MNRAS.292...27H}) and assuming all ionized gas is
contained in fully ionized bubbles. We integrate the LF down to
$M_\mathrm{AB}(UV) = -13.04$, equivalent to a star formation rate of
$0.01~m_{\sun}/\textrm{yr}$ \citep{1998ARA&A..36..189K,
  2009ApJ...690.1350O}. We use the ``WMAP7 + BAO + H0'' mean
cosmological parameters from \citet{arXiv:1001.4538} throughout this
paper. The values of $f_{\mathrm{esc}\gamma}$, $\beta_{M*}$, and
$\alpha$ are discussed in \S \ref{sec_results}.

\subsection{Connecting Reionization and Carbon Production}\label{sec_fxCIV}

We define $f_{xCIV}$ as the ratio between the total rate (with no
delay) of \ion{C}{iv} production and the total rate of ionizing photon
production, so that:
\begin{equation}
 \OmegaCIVInstantDot(z) = \dot{x}_\mathrm{total}(z) f_{xCIV} \textrm{,}
\end{equation}
where $\OmegaCIVInstantDot(z)$ would be the rate of \ion{C}{iv}
density increase with no delay in emission, and
$\dot{x}_\mathrm{total}(z)$ is the ratio of the total rate density of
ionizing photon production to the total number density of hydrogen and
helium atoms.

The $f_{xCIV}$ ratio depends on a number of factors:
\begin{equation}
f_{xCIV} = \left[\frac{f_\mathrm{CIV} f_{\mathrm{esc}Z}}{r_{\gamma Z}}\right] (X_C/Z) \Omega_\mathrm{baryon}\textrm{.}
\end{equation}
The fraction of all metal mass in carbon is $X_C/Z = 0.178$
\citep*[the solar value from][]{2005ASPC..336...25A}.%
$\Omega_\mathrm{baryon} = 0.0456$ is the fraction of the critical
density contributed by baryons \citep{arXiv:1001.4538}. The terms in
square brackets (the fraction of carbon in \ion{C}{iv},
$f_\mathrm{CIV}$; the fraction of metals that escape galaxies,
$f_{\mathrm{esc}Z}$; and the ratio of metal nucleon to ionizing photon
production, $r_{\gamma Z}$) are highly uncertain. We use the
representative values of $f_\mathrm{CIV}=0.5$ \citep[the maximum
  theoretical value, see][]{2009MNRAS.395.1476R,2001ApJ...561L.153S}
and $f_{\mathrm{esc}Z} = 0.2$ (equal to our fiducial
$f_{\mathrm{esc}\gamma}$), and emphasise that it is only the product
of these uncertain factors that matters.

The ratio of ionizing photons to metal nucleons produced by a stellar
population can be expressed as
\begin{equation}
r_{\gamma Z} = \eta \frac{E_\mathrm{p}}{E_\mathrm{avg}}\textrm{,}
\end{equation}
where $\eta$ is the ratio of total ionizing photon energy to total
rest--mass energy of the metals produced in a stellar population,
$E_\mathrm{p}$ is the rest--mass energy of a proton, and
$E_\mathrm{avg}$ is mean energy of ionizing photons
\citep{2002A&A...382...28S}. For a stellar population with $Z = 1/50
Z_{\sun}$, \citet{2002A&A...382...28S} calculate $\eta=0.014$ and
$E_\mathrm{avg} = 21.95$~eV, yielding $f_{xCIV} = 1.4 \times
10^{-9}$. Note that this $\eta$ value does not include yields from
low-- and intermediate--mass stars (LIMS, $m < 8 m_{\sun}$), stellar
wind mass loss, or Type I SN contributions, all of which would
decrease $\eta$ (and increase the enrichment to ionization ratio
$f_{xCIV}$). We will also explore scenarios using the
solar-metallicity ($Z=Z_{\sun}$) values of $\eta=0.0036$ and
$E_\mathrm{avg}=20.84$~eV (calculated with mass loss and SN Ibc, but
still without LIMS), resulting in $f_{xCIV} = 5.0 \times 10^{-9}$.

Changing the $M_*$ vs. $z$ slope ($\beta_{M*}$) or the faint--end slope
of the LF ($\alpha$) changes both the ionization and enrichment
histories. Changing $f_{\mathrm{esc}\gamma}$ only affects the
ionization history. Changing $f_{xCIV}$ only affects the enrichment
history.

\subsection{The Carbon Delay Distribution Due to Stellar Lifetimes}\label{sec_delay}

\begin{figure}
\includegraphics[width=3in]{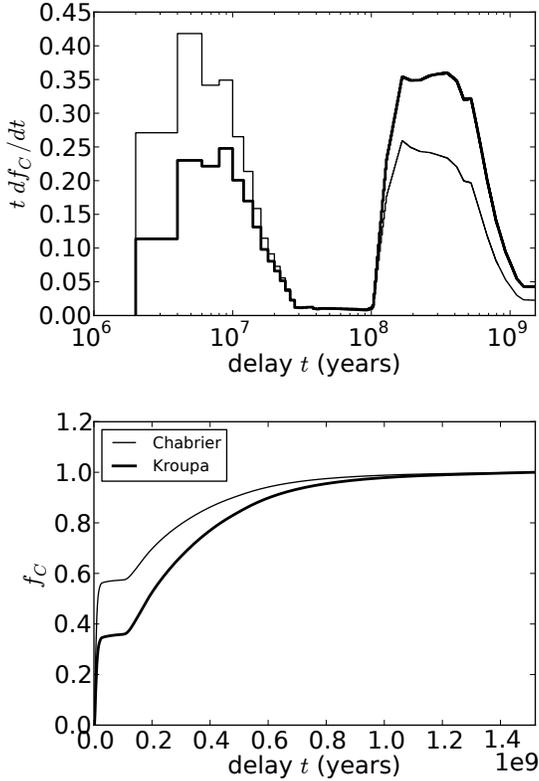}
\caption{\label{fig_delay} Delay functions for carbon
  production. Differential (top panel) and cumulative (bottom panel),
  delays are calculated with Chabrier (thin curve) and Kroupa (thick
  curve) IMFs. The differential delay is shown on a logarithmic time
  scale, while the cumulative delay is shown on a linear time
  scale. Note the rapid emission from supernovae ($\sim 10^7$~years)
  and the longer tail from AGB stars ($\sim$ a few $\times
  10^8$~years).}
\end{figure}

Carbon is not produced instantly upon formation of a population of
stars, unlike, for our purposes, ionizing photons. Instead carbon is
ejected primarily after the main--sequence lifetime of a star is over,
which for low and intermediate mass stars ($m \la 8 M_{\sun}$,
$t_\mathrm{life} \ga 3 \times 10^7$ years) becomes a significant
fraction of the relevant timescales (e.g. the $3 \times 10^8$~year
interval from $z=6$ to $5$).

Neglecting this delay \citep[as in][]{arXiv:1005.1451} is often
justified with the statement that Type II supernova (SNII) yields from
short--lived, high--mass stars dominate carbon production at high
redshift. However, even if most most carbon is synthesised in SNeII%
\footnote{The fraction of carbon contributed by low and intermediate
  mass stars depends sensitively on the amount of ``hot bottom
  burning'' (HBB) that takes place on the asymptotic giant branch
  (AGB), since HBB can destroy carbon and even result in a net loss of
  $^{12}C$ over the lifetime of a star. The amount of HBB as a
  function of stellar mass is still highly uncertain,
  \citep{arXiv:1007.2533} so it is unclear whether low-- or high--mass
  stars dominate $^{12}C$ production \citep[][and references
    therein]{arXiv:1006.5863}.}
, a substantial fraction of that carbon is incorporated into low-- and
intermediate--mass stars (LIMS, $m \la 8 m_{\sun}$) before being blown
into the IGM, according to the models of
\citet{2008MNRAS.387..577O}. In other words, an important fraction of
carbon produced in a galaxy gets locked up in LIMS and is only
returned to the gas phase (and made available for ejection in a
galactic outflow) during the asymptotic giant branch (AGB) phase,
after the main--sequence lifetime has elapsed. Therefore, the
lifetimes of lower--mass stars impose a delay on the ejection of some
of the carbon into the IGM, and AGB star ejection of carbon cannot be
neglected in calculating the timing of IGM enrichment.

Since we are trying to model the connection between ionizing emission
associated with star formation and the eventual enrichment of the IGM
with carbon produced by the same stars, it is important that we take
this delay into account. Indeed, such a delay is inevitable, and the
original motivation of this paper was to assess whether this delay
might help explain the steep observed evolution of the \ion{C}{iv}
abundance in the IGM, as discussed in \S \ref{sec_introduction}.

The cumulative delay function $f_C(t)$ is the fraction of carbon
emitted by stars with lifetimes $t_\mathrm{life} < t$. Assuming all
carbon is ejected at the end of a star's main sequence lifetime, the
fraction of carbon ejected in the time interval $t \pm (dt/2)$ after
star formation is
\begin{equation}
  \frac{d f_C}{dt}(t) = \frac{d f_C}{dm} \frac{dm}{dt}\textrm{,}
\end{equation}
where $dm/dt$ is the inverse of the derivative of the lifetime
function ($t_\mathrm{life}(m)$). The fraction of carbon produced by
stars of mass $m \pm (dm/2)$ is
\begin{equation}
  \frac{d f_C}{dm}(m) \propto M_C(m) \phi(m)\textrm{,}
\end{equation}
where $M_C(m)$ is the carbon mass ejected by a star of mass $m$, and
$\phi(m)$ is the initial mass function (IMF) of stars by number. We
normalise this function to give
$\int_{t_\mathrm{min}}^{t_\mathrm{max}} d f_C/dt = 1$. Here
$t_\mathrm{min} = 3.24 \times 10^6$ years is the lifetime of a $100
M_{\sun}$ star, and $t_\mathrm{max} = 1.52 \times 10^9$ years is the
cosmic time between the starting and ending points of our simulations,
$z=100$ and $z=4.1$. With our chosen IMFs and yields (see below), an
additional $< 9\%$ of carbon would emerge at $t > t_\mathrm{max}$.

Convolving the $\OmegaCIVInstant(t)$ curve
with $\frac{d f_C}{dt}$ gives us the delayed \ion{C}{iv} curve
\begin{equation}
  \OmegaCIVDelay(t) = \OmegaCIVInstant * \left (\frac{d f_C}{dt}\right
  ),
\end{equation}
where $*$ represents convolution over the time coordinate.

\citet{2005A&A...430..491R, arXiv:1006.5863} have quantified in detail
the impact of uncertainties in the IMF, stellar lifetimes, and stellar
yields on Galactic chemical evolution models. The greatest
uncertainties are in the yield calculations, which vary considerably
from author to author. As \citet{arXiv:1006.5863} point out, there is
no consistent set of yields covering the whole range of mass and
metallicity and including all of the physical effects relevant to
either galactic or cosmic chemical evolution models, and essentially
no suitable calculations have been performed for $m \approx 6$ -- $8
M_{\sun}$. For convenience, we use a set of carbon ($^{12}$C) yield
values provided by \citet*{2005A&A...432..861G}%
\footnote{Available on VizieR: \url{http://vizier.u-strasbg.fr}}
, containing their own original calculations for $m = 1$--$8
M_{\sun}$, and \citet{1995ApJS..101..181W} values for $m = 8$ -- $100
M_{\sun}$. We use the yields for metallicity $Z = 0.02$ because the
variation in calculated carbon yield with metallicity (at least above
some threshold) is smaller than the overall uncertainty in yields.

The final complication with the use of yield tables is the distinction
between the total mass of an element ejected by a star and the net
yield of new atoms synthesised in the star
\citep{2005A&A...432..861G}. In a self--consistent chemical evolution
model that tracks the metallicity of the star--forming environment,
the yield of new elements is the relevant quantity. However, we are
only tracking carbon abundance in the IGM, while stars are forming
directly out of gas in the interstellar medium (ISM), so we use the
total ejected mass of carbon to calculate the delay. Note that this is
independent of the calculation of the total amount of carbon produced
(see \S \ref{sec_fxCIV}). This is a good approximation if the ISM
reaches a stable metallicity quickly, and the composition of the
galactic outflow is representative of the total mass currently being
ejected from stars (both via SN and AGB mass loss).

Theoretical calculations of stellar lifetimes generally agree fairly
well for $m > 1 M_{\sun}$. The dependence on metallicity is quite
weak. The larger uncertainties at $m < 1 M_{\sun}$ are irrelevant here
since the corresponding lifetime of $t_\mathrm{life} >~ 10$~Gyr is
longer than the age of the universe at $z >~ 0.3$. We adopt the
lifetime function of Kodama as given in \citet{2005A&A...430..491R}.

The stellar initial mass function is another important source of
uncertainty. We can characterise the impact of the IMF on the delay
function by dividing the carbon emission into a prompt component and a
delayed component. We are concerned here with evolution on a timescale
of $\sim 10^8$ years. Therefore carbon emission that occurs faster
than $10^7$ years after star formation is relatively prompt. Of the
IMFs discussed in the \citet{2005A&A...430..491R} review, the
\citet*{1993MNRAS.262..545K} and \citet{2003PASP..115..763C} IMFs
produce the most extreme values for the cumulative delay function at
$10^7$ years, $f_C(10^7 \mathrm{yr})$. With a Kroupa IMF $f_C(10^7
\mathrm{yr})=0.23$, while $f_C(10^7 \mathrm{yr})=0.44$ for a Chabrier
IMF. We therefore calculate all of our models with both of these IMFs%
\footnote{Note that we are only varying the IMF in the calculation of
  the delay function. In principal $\dot{N}/L(1500~\textrm{\AA})$ and
  $r_{\gamma Z}$ also depend on the IMF and metallicity of the stellar
  population, but we treat each of these calculations independently.}
in order to demonstrate quantitatively the impact of the IMF
uncertainty on our results, and to suggest qualitatively the impact
that different sets of yield values might have. Figure \ref{fig_delay}
shows the differential and cumulative delay functions with Kroupa and
Chabrier IMFs.  These figures illustrate that roughly half of the
carbon ejection occurs essentially instantly, whereas the remaining
half is spread over the lifetime ($\sim 1$Gyr) of intermediate--mass
stars.

Extremely low metallicities may result in dramatically different IMFs
and yields from those assumed here. We do not consider this
metal--free, or Population--III, mode of star formation in calculating
the delay function, as metal--free stars are thought to make up only a
small fraction of the stars formed before $z=6$, even if their
formation continues at a low rates to late times \cite[see,
  e.g.][]{2009MNRAS.398.1782R,arXiv:1003.3873}.

\section{Results}\label{sec_results}

The primary constraints on the epoch of reionization available today
are the {\it WMAP} measurement of $\tau$ (the optical depth to
Thompson scattering from free electrons in the IGM,
\citealp{arXiv:1001.4538}) and the evolving Lyman--$\alpha$ opacity of
the IGM at $z \sim 6$ (though see \citealp{2010MNRAS.tmp.1017M} for a
discussion of the complicated, model--dependent interpretation of this
evolution). In \S \ref{sec_reionresults}, we discuss the agreement
between our luminosity function histories and these reionization
constrains. Then in sections \ref{sec_CIVresults} and
\ref{sec_riseresults}, using the formalism outlined above, we return
to the puzzles posed in the introduction.

\subsection{Matching Reionization Constraints}\label{sec_reionresults}

In this section, we compare our luminosity function histories to
available constraints on the reionization of the IGM. For our fiducial
luminosity function history, we adopt a Schechter luminosity function
with fixed $\phi_*=1.1\times{}10^{-3}~\mathrm{Mpc}^{-3}$ and $\alpha$,
and use the observed $M_*$ values from \citet{2008ApJ...686..230B}
(linearly interpolated) from $z=3.8$ to $9.0$. We extrapolated $M_*$
linearly in $z$ above $z=9$ with slope $\beta_{M*}$. The extrapolation
to lower luminosities is controlled by the faint--end slope,
$\alpha$. \citet{2009ApJ...690.1350O} suggest that $M_*$ is
approximately linear in $z$ at high redshift, with a slope of
$\beta_{M*} = 0.36 \pm 0.18$. \citet{2008ApJ...686..230B} suggest
$\alpha=-1.74$ at high redshift. We use $f_{\mathrm{esc}\gamma} =
0.2$.

\begin{figure}
\includegraphics[width=3in]{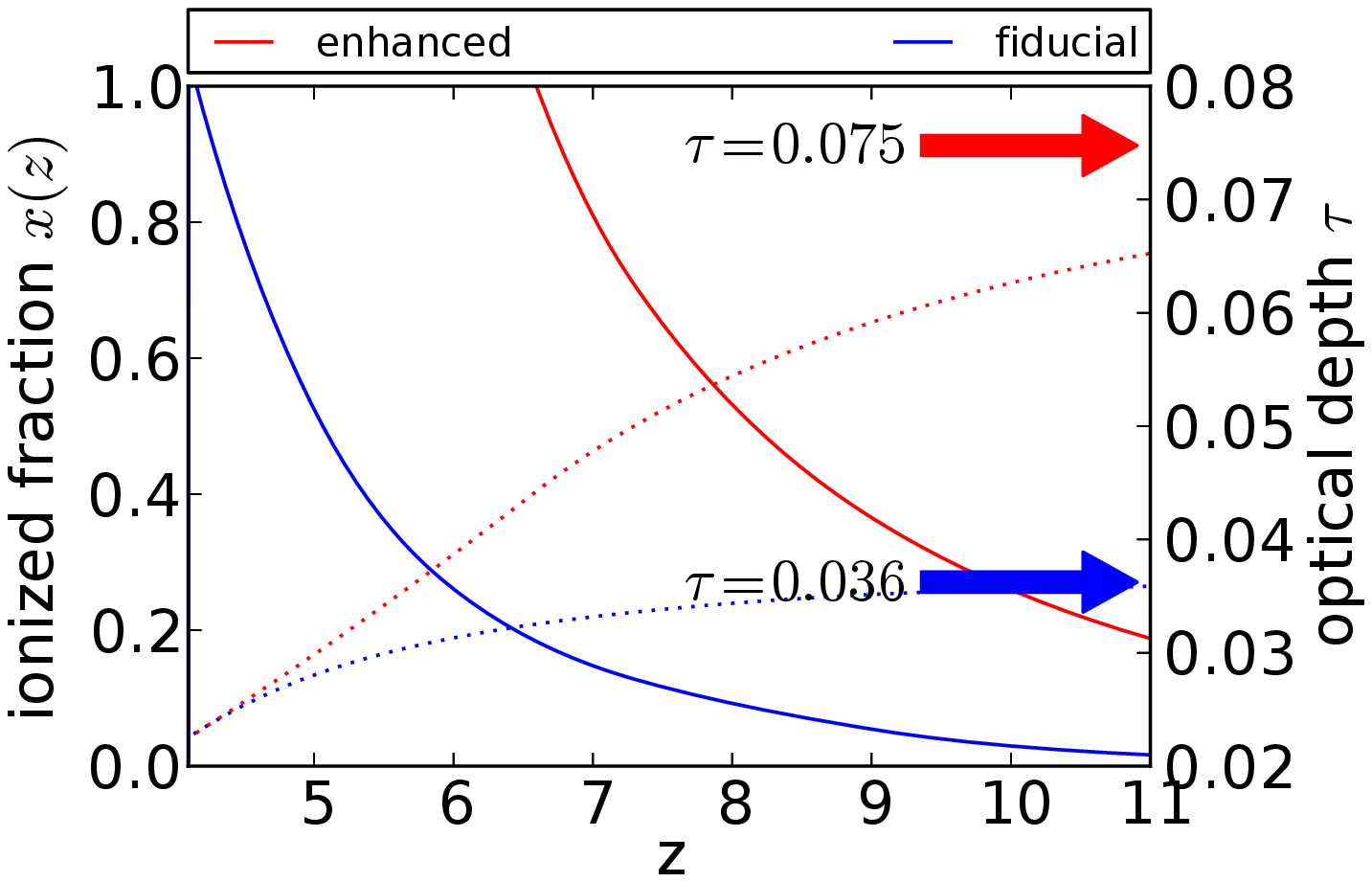}\\
\includegraphics[width=3in]{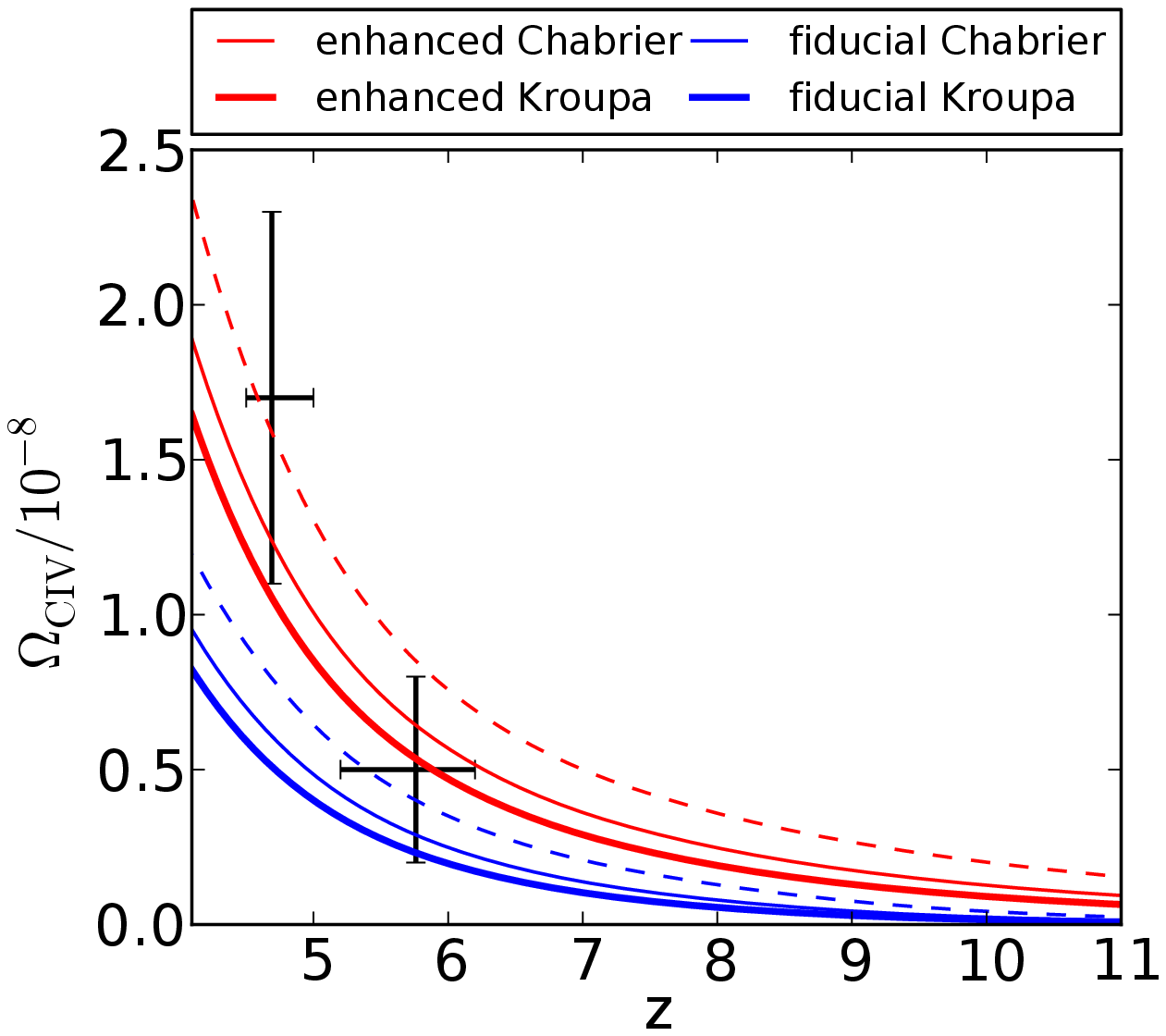}\\
\includegraphics[width=3in]{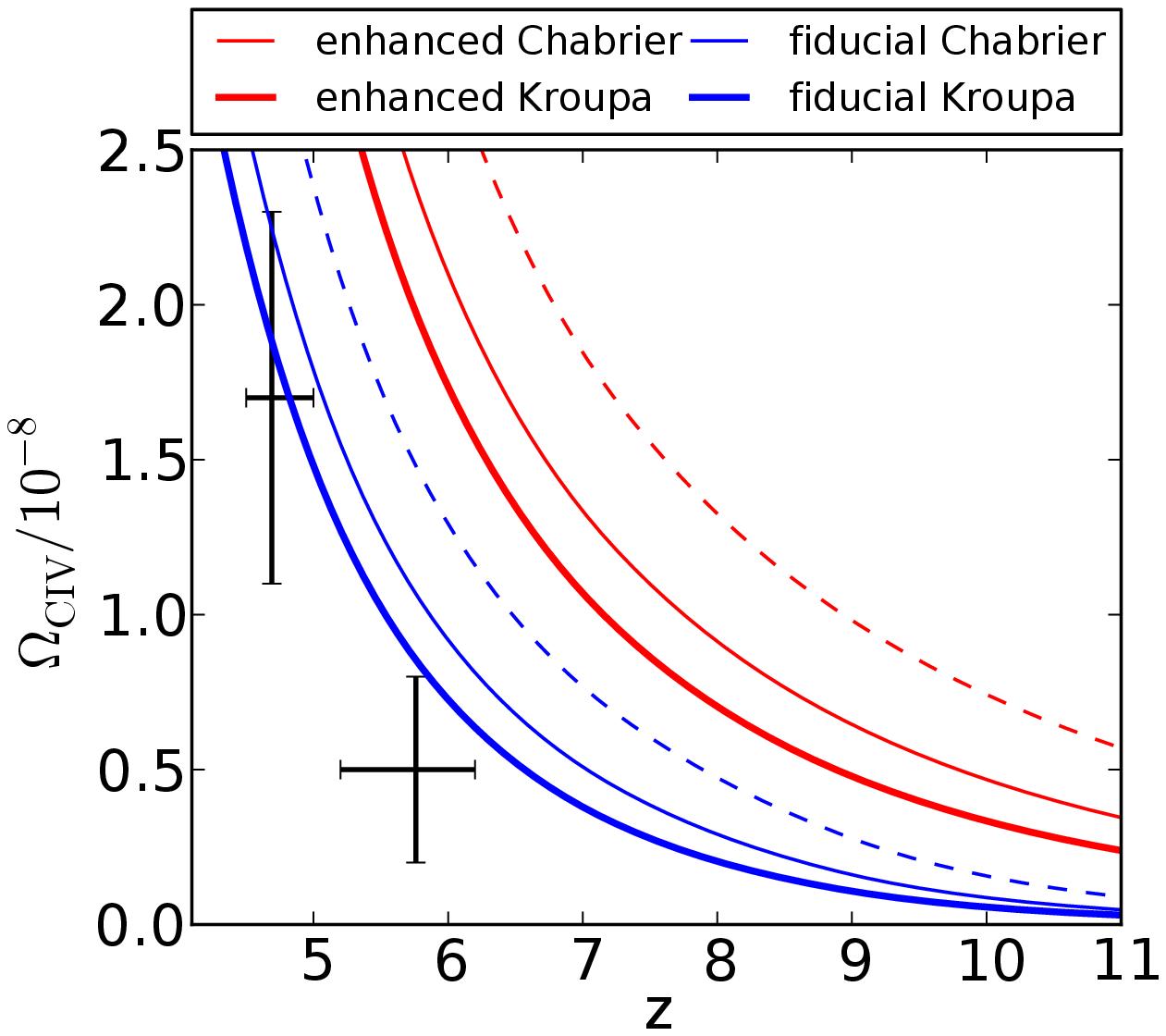}
\caption{\label{fig_ionenrich} Ionization and enrichment histories of
  the IGM. In each panel, the lower (blue) set of curves corresponds
  to the fiducial LF history, and the upper (red) set to the enhanced
  LF chosen to match {\it WMAP} $\tau=0.087$ and $z_\mathrm{reion}
  \sim 6.6$. Top panel: ionized fraction $x(z)$ (solid) and optical
  depth $\tau(<z)$ (dotted). The arrows on the right show the total
  value of the optical depth (integrated to $z=100$). Middle panel:
  \ion{C}{iv} density in the IGM with (solid) and without (dashed) a
  delay function. The thin solid curve uses the Chabrier--IMF delay,
  while the thick curve uses the Kroupa--IMF delay. Observed values
  are indicated as black data points with error bars. Horizontal error
  bars indicate the interval over which the density is averaged, and
  vertical error bars indicate the uncertainty. Bottom panel:
  Enrichment in the same models as in the middle panel, except with
  $f_{xCIV}$ increased by a factor of $3.7$. This increase is
  equivalent to the difference between $1/50$--solar and
  solar--metallicity values for the ratio of ionizing photons to metal
  nucleons.}
\end{figure}

Figure \ref{fig_ionenrich} shows the ionization and enrichment
histories as functions of redshift. The lower (blue) set of curves
corresponds to the fiducial parameters described above. The top panel
shows the ionized fraction in the IGM, along with the optical depth
integrated from redshift $0$ to $z$. The arrows on the right show the
total value of the optical depth (integrated to $z=100$). As
\citet{2009ApJ...690.1350O} have pointed out \citep[see
  also][]{2007MNRAS.382..325B}, combining this luminosity function
evolution with reasonable ionizing photon escape fraction
($f_{\mathrm{esc}\gamma} = 0.2$) and IGM clumping factor ($C = 4$)
values, yields an insufficient emissivity to either complete
reionization by $z \sim 6$ or match the {\it WMAP} constraint on the
electron scattering optical depth \citep[$\tau = 0.087 \pm
  0.014$][]{arXiv:1001.4538}. With the fiducial LF history, the IGM is
only $20\%$ ionized by $z=6$, and the optical depth is $\tau=0.036$.

In order to match the {\it WMAP} $\tau$ value and complete
reionization at $z \ga 6$, we increase $f_{\mathrm{esc}\gamma}
\dot{N}/L(1500~\textrm{\AA})$ by a factor of $3$, flatten the $M_*$
slope to $\beta_{M*}=0.09$, and set the faint end slope to the steeper
value of $\alpha=-1.95$. The higher $f_{\mathrm{esc}\gamma}
\dot{N}/L(1500~\textrm{\AA})$ could be explained by a higher escape
fraction, or a higher $\dot{N}/L(1500~\textrm{\AA})$ due to a lower
metallicity or more top-heavy IMF of the stellar population. For
instance, \citet{2008ApJ...680...32C} finds a factor of $\sim 2$--$3$
increase in $\dot{N}/L(1500~\textrm{\AA})$ when the metallicity falls
from $0.4 Z_{\sun}$ to $0.02 Z_{\sun}$. Note that such changes would
affect the delay function and $r_{\gamma Z}$ values as well.  The
changes to $\beta_{M*}$ and $\alpha$ make the enhanced LF history
resemble the recent results by \citet{arXiv:1006.4360}, who found a
brighter $M_*$ and steeper $\alpha$ at $z=7$ and $8$ than were
suggested by earlier results. We integrate the LF down to
$M_\mathrm{AB}(UV) = -13.04$, equivalent to a star formation rate of
$0.01~m_{\sun}/\textrm{yr}$ \citep{1998ARA&A..36..189K,
  2009ApJ...690.1350O}. Adjusting any of these parameters alone cannot
match both constraints, and adjusting them simultaneously allows us to
use more plausible values. More important than the exact parameter
values is the resulting ionizing emissivity history. This is obviously
not a unique solution, but we present this enhanced LF history as a
plausible example of one that matches current observational
constraints much better than the fiducial extrapolation of the
observed LBG luminosity function. The upper (red) sets of curves in
Figure \ref{fig_ionenrich} correspond to this enhanced LF history. The
top panel shows that the optical depth has been increased to near the
{\it WMAP} value, and the IGM is fully ionized by $z=6.6$. Table
\ref{tab_LFHparams} summarises the parameters of each LF history.

We use a constant clumping factor $C=4$ in these calculations, though
our basic conclusion that the enhanced LF history is consistent with
existing reionization constraints is not particularly sensitive to
changes in this assumption. For instance, using a higher constant
$C=6$ results in $\tau = 0.067$, $z_\mathrm{reion} = 6.21$ for the
enhanced LF history. The clumping factor should actually be lower at
higher redshift, however. \citet{2008ApJ...680...32C} has derived the
clumping factor as a function of redshift for the relevant gas
(ionized gas outside of ionizing--photon source halos) from
simulations by \citet{2007ApJ...671....1T}. Using their clumping
factor history (estimated from their Figure 2a), results in a larger
$\tau = 0.082$ (because the clumping factor is lower, $C<4$, at early
times, $z>10.5$) and a later $z_\mathrm{reion}=5.7$ (because $C \ga 8$
at $z \la 7$). This optical depth is quite close to the WMAP7
value. \citet*{arXiv:0912.3034, 2009MNRAS.394.1812P} found that
reheating of the IGM results in an even lower clumping factor history,
which would require less enhancement in the LF history (something
between our ``fiducial'' and ``enhanced'' LFs) to match reionization
constraints.

\begin{table}
\caption{Parameters of our luminosity function (LF) histories.}
\label{tab_LFHparams}
\begin{tabular}{lrrrrl}
\hline
name & $\alpha$ & $\beta_{M*}$ & $f_{\mathrm{esc}\gamma}$ &
$z_\mathrm{reion}$ & $\tau$ \\ 
\hline
fiducial & -1.74 & 0.36 & 0.2 & 4.2 & 0.036\\
enhanced & -1.95 & 0.09 & 0.6 & 6.6 & 0.075\medskip\\
\multicolumn{5}{r}{\textit{WMAP7 \citep{arXiv:1001.4538}:}} & $0.087 \pm
  0.014$\\
\hline
\end{tabular}\medskip

\textbf{Parameters:}\\
$\alpha$ is the faint--end slope of the Schechter luminosity
function.

$\beta_{M*}$ is the slope of $M_*$ as a function of $z$.

$f_{\mathrm{esc}\gamma}$ is the escape fraction of ionizing photons
from galaxies. Note that it is completely degenerate with the SED
slope, which we fix at $\dot{N}/L(1500~\textrm{\AA}) = 8.4\times
10^{24}~\textrm{photons}~\textrm{s}^{-1}/(\textrm{erg}~\textrm{s}^{-1}~\textrm{Hz}^{-1})$.

\textbf{Results:}\\
$z_\mathrm{reion}$ is the redshift at which the ionized fraction $x(z)
= 1$.

$\tau$ is the optical depth due to free electrons.

\end{table}

\subsection{Matching CIV Abundance}\label{sec_CIVresults}

The bottom two panels of Figure \ref{fig_ionenrich} display the
\ion{C}{iv} mass density as a fraction of the critical density
$\OmegaCIV(z)$. The dashed lines assume instantaneous ejection, the
thin solid lines are convolved with the Chabrier--IMF--based delay
function, and the thick solid lines with the longer Kroupa--IMF
delay. 

In order to show the effect of the large uncertainty in the value of
$f_{xCIV}$, in the middle panel we use the $1/50~Z_{\sun}$ value for
$f_{xCIV}$, while in the bottom panel we use the solar--metallicity
value (see \S \ref{sec_fxCIV}). The solar--metallicity stellar
population produces fewer ionizing photons per metal nucleon
synthesised, so for a fixed LF history it produces higher \OmegaCIV{}
values. Both of the puzzles discussed in \S \ref{sec_introduction} are
evident in the middle panel, which effectively uses the same
$f_{xCIV}$ assumed by \citet{2009MNRAS.395.1476R}.

First we can see the conflict suggested by \citet{2009MNRAS.395.1476R}
between reionization constraints and the low $z=5.8$ $\OmegaCIV$ value
in the middle panel of Figure \ref{fig_ionenrich}. The enrichment
history calculated from the LF history that matches reionization
constraints (upper/red dashed curve) produces too much \ion{C}{iv} at
$z = 5.8$.%
\footnote{Again, we are ignoring the large errors on the observations
  for the sake of exploring their consequences should they prove to be
  accurate. If we take the observational errors into account, then we
  can see that even the upper/red dashed curve is marginally
  consistent with the observations.}  This overproduction at a single
redshift, in and of itself, is not too troubling. Either a slight
decrease in the highly--uncertain $f_{xCIV}$ or a slight decrease in
the star formation rate density (SFRD) could lower $\OmegaCIV$
sufficiently to agree with the $z=5.8$ measurement. Also, the
stellar--lifetime delay, especially with the less top--heavy Kroupa
IMF, brings $\OmegaCIV$ down to close to the observed value.

The second puzzle --- the rapid buildup of \ion{C}{iv} from $z=5.8$ to
$4.7$ noted by \citet{2009ApJ...698.1010B} --- is also evident in
Figure \ref{fig_ionenrich}. In the middle panel, again, the enhanced
LF curve with no delay is close to matching the $z=4.7$ observation,
but slightly over-predicts the earlier $z=5.8$ point. Similarly, the
fiducial LF curve with no delay only slightly under-predicts the
$z=5.8$ value, but is much lower than the later $z=4.7$
observation. The predicted evolution of \OmegaCIV{} using either LF
history is too slow to match the observations.

We were motivated to calculate the stellar--lifetime--based enrichment
delay by the idea that such a delay might help to explain to rapid
rise in \ion{C}{iv}. In fact, as Figure \ref{fig_ionenrich} clearly
illustrates, we find that stellar lifetimes contribute little to the
solution of this puzzle. In the next subsection, we explore the effect
of longer delays.

\subsection{The Rapid Rise in CIV}\label{sec_riseresults}

Since $f_{xCIV}$ is so uncertain, it is useful to find a quantity
independent of $f_{xCIV}$ to compare with the observations. For a
given combination of an ionizing emissivity history and a delay
function, the fractional increase in \OmegaCIV{} over a specific
redshift interval is fixed and does not depend on $f_{xCIV}$ (as long
as $f_{xCIV}$ is independent of redshift). The observed fractional
increase is
\begin{equation}
\frac{\Delta \OmegaCIV}{\OmegaCIV} \equiv
\frac{\OmegaCIV(4.7)}{\OmegaCIV(5.8)} - 1 = 2.4 \textrm{.}
\end{equation}

To compare this number with our theoretical curves, we average
\OmegaCIV{} over the same intervals used to determine the observed
values. With no delay, the fractional increase is $\Delta
\OmegaCIV/\OmegaCIV = 0.79$ for the fiducial LF history, and $0.70$
for the enhanced LF history. These values are far smaller than the
observed increase.

We expect a delay in enrichment to make the fractional increase in
\ion{C}{iv} larger, because (if \OmegaCIV{} evolves slower than
exponentially) the fractional rate of increase at earlier times must
be higher. To see this, consider that the fractional increase is
roughly
\begin{equation}
\frac{\Delta \OmegaCIV}{\OmegaCIV} \approx \frac{\OmegaCIVDot}{\OmegaCIV}(t) \Delta t,
\end{equation}
where $\Delta t \sim 0.3$ Gyr is the time interval from $z=5.8$--$4.7$
and $\OmegaCIVDot$ is the derivative of $\OmegaCIV$ with respect to
cosmic time. If \OmegaCIV{} is a power--law, proportional to $t^n$ 
(this is a reasonable approximation at the cosmic times we are considering, 
and it is also a conservative one, in the sense that structure formation 
is increasingly more rapid at higher redshifts),
then 
\begin{equation}
\frac{\OmegaCIVDot}{\OmegaCIV}(t) = \frac{n}{t}\textrm{.}
\end{equation}
Therefore, the fractional growth rate decreases with time. If we had a
delta--function delay, so that $\OmegaCIVDelay(t) = \OmegaCIVInstant(t
- t_\mathrm{delay})$, then the delay would boost the fractional
increase by a factor of
\begin{equation} \label{eq_delay_boost}
 \frac{\OmegaCIVDelayDot/\OmegaCIVDelay}{\OmegaCIVInstantDot/\OmegaCIVInstant}
\sim \frac{t}{t - t_\mathrm{delay}}.
\end{equation}
We need a factor of $t/(t-t_\mathrm{delay}) = 3$--$3.5$ to boost the
fractional \ion{C}{iv} increase from $\Delta \OmegaCIV / \OmegaCIV =
0.7$--$0.8$ to $2.4$. We can conclude that a delta--function delay of
$\sim 0.7$~Gyr should boost the growth sufficiently to match the slope
inferred from the two observations.  Our stellar--lifetime--based
delays, however, are too short to provide the necessary boost. The
mean delay with the Chabrier IMF is $0.16$ Gyr. With the Kroupa IMF,
the mean delay is only slightly longer, $0.25$ Gyr. Furthermore, and
even more problematically, the delay is not a delta function. A large
fraction (roughly half; see Figure~\ref{fig_delay}) of the carbon is
ejected promptly, which dilutes the boosting effect.

\begin{figure}
\includegraphics[width=3in]{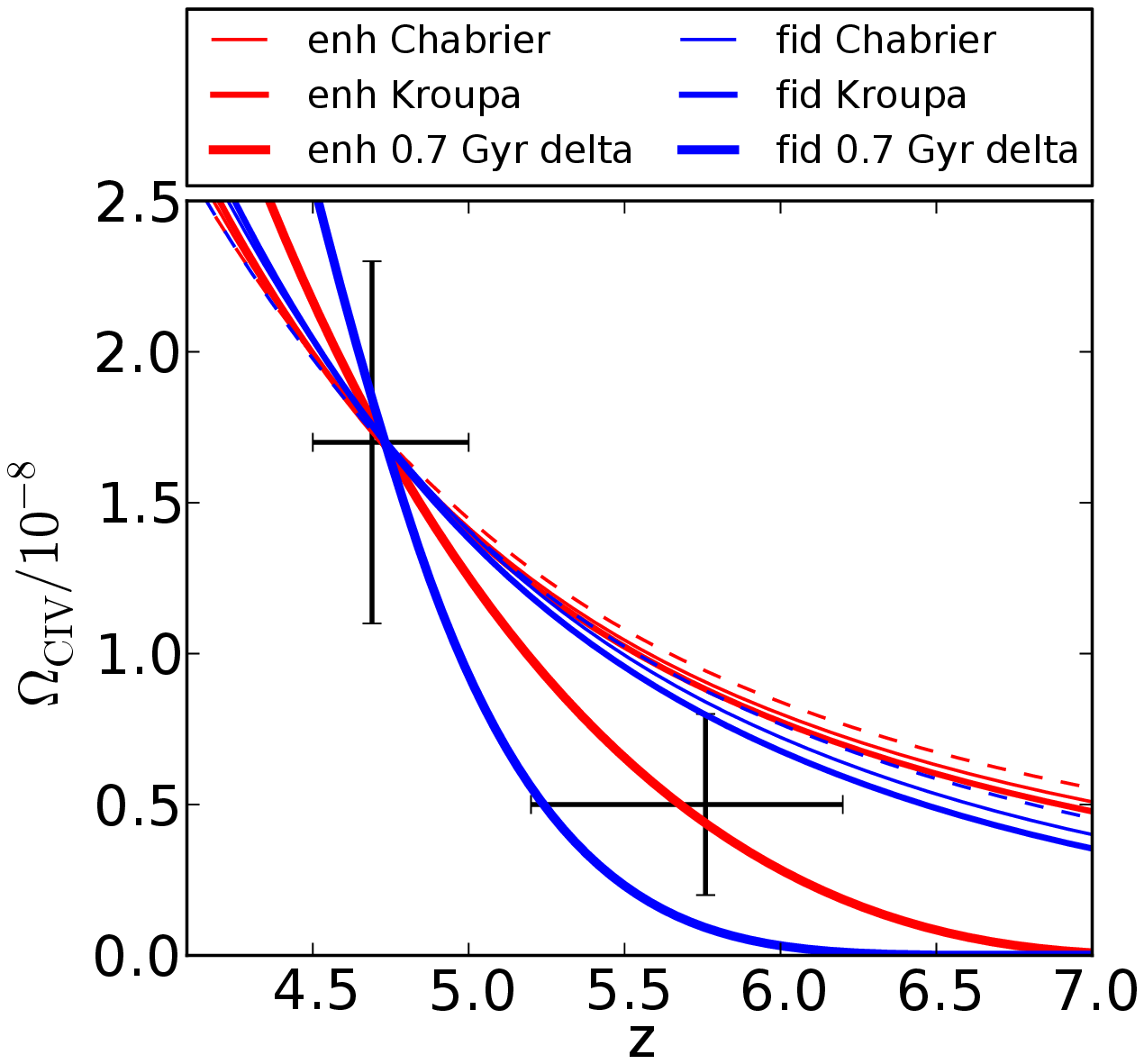}\\
\includegraphics[width=3in]{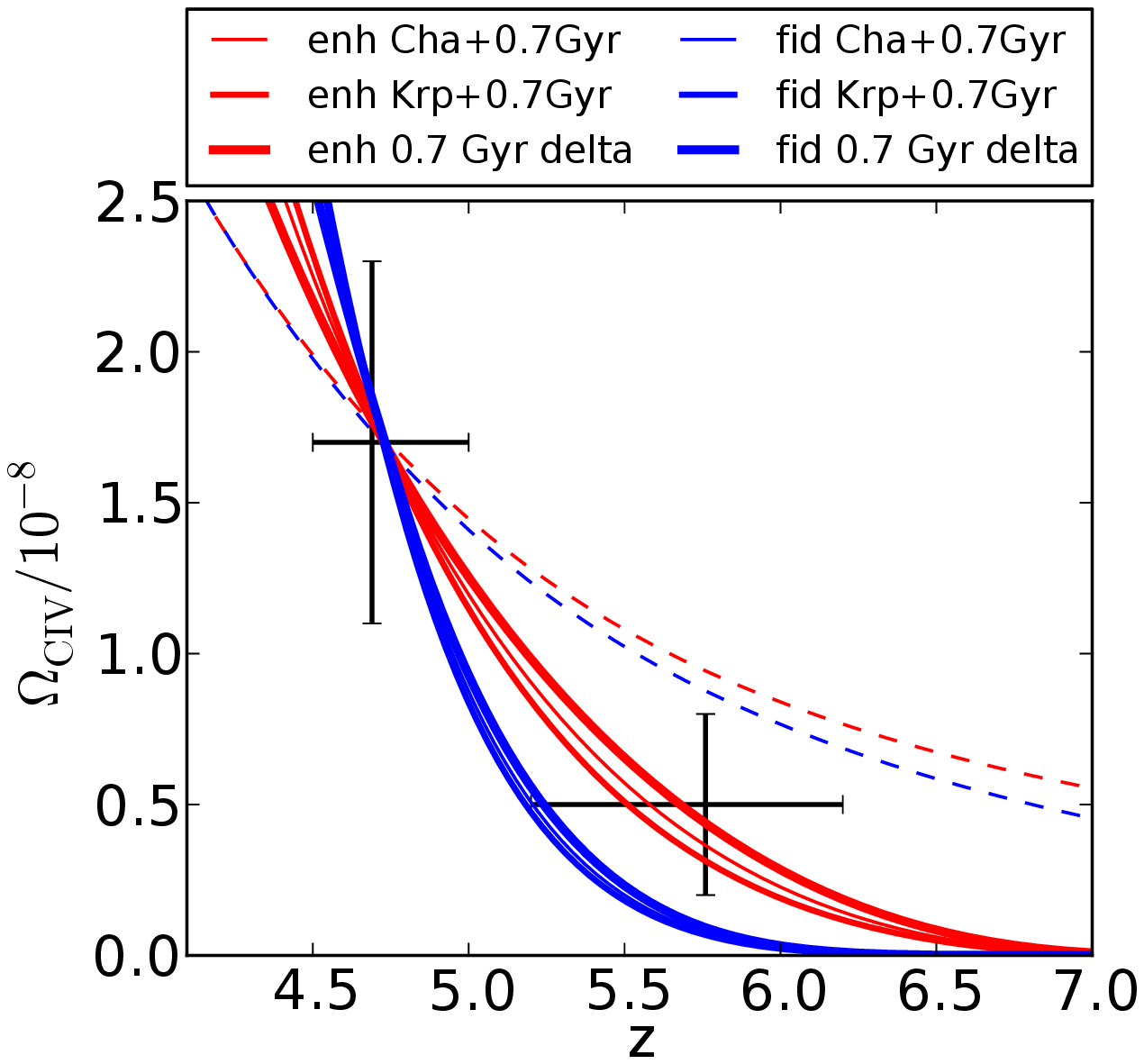}\\
\includegraphics[width=3in]{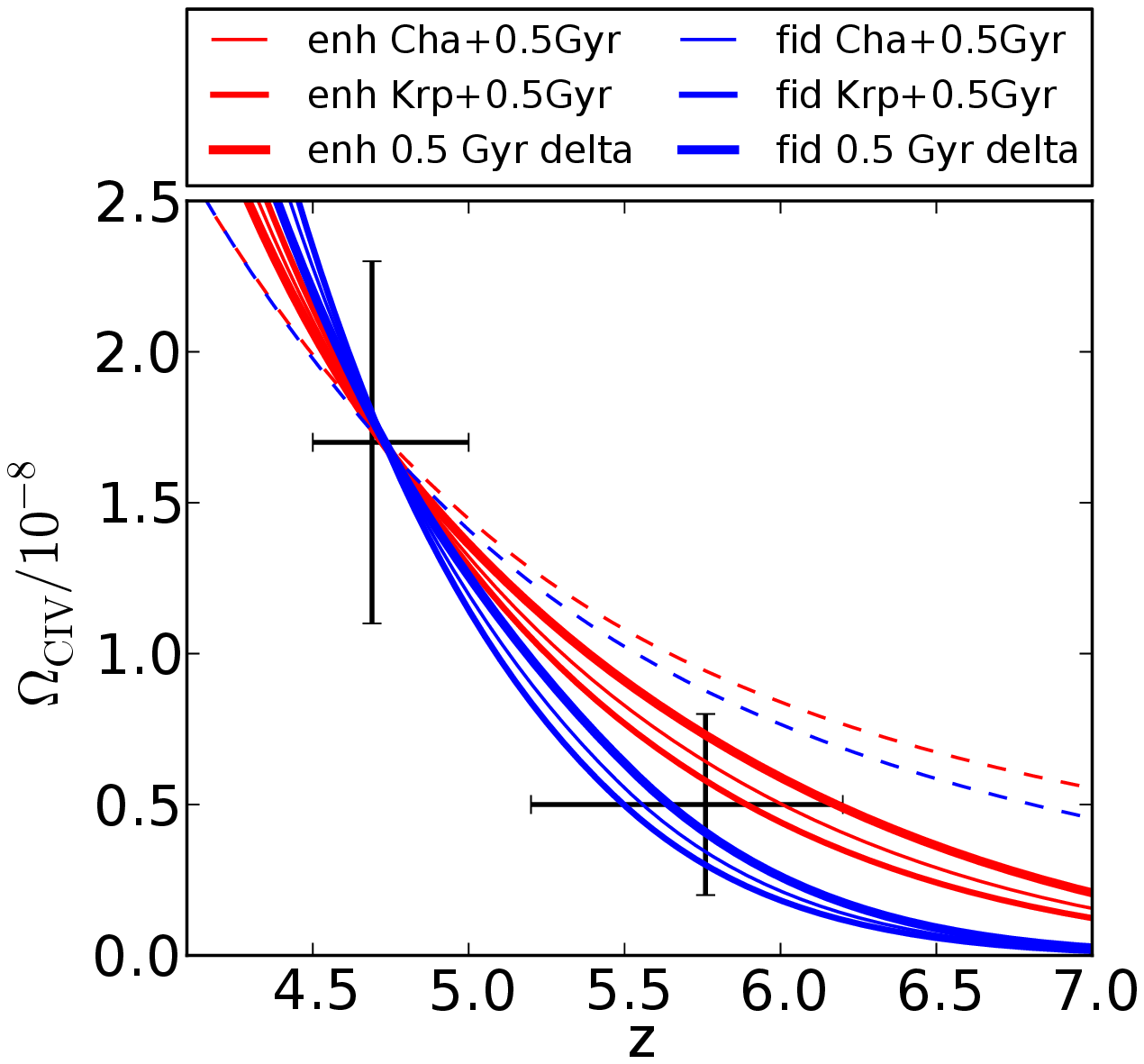}
\caption{\label{fig_delayedenrich} Enrichment histories with the
  stellar carbon--to--ionizing--photon production ratio, $f_{xCIV}$,
  adjusted to match the $z=4.7$ observation. The dashed curves assume
  instantaneous production of \ion{C}{iv}. Top panel: comparing
  stellar--lifetime--based delays and a $0.7$~Gyr delta function
  delay. Middle panel: a $0.7$~Gyr delay has been added to the
  stellar--lifetime delays. Bottom panel: a $0.5$~Gyr delay has been
  added to stellar--lifetime delays.}
\end{figure}

\begin{table}
\caption{Fractional increase in \ion{C}{iv} density for each
  luminosity function history and delay function combination.}
\label{tab_increase}
\begin{tabular}{llrr}
\hline
LF History & Delay Function & $f_{xCIV}/10^{-9}$ &
$\Delta\OmegaCIV{}/\OmegaCIV$\\
\multicolumn{2}{c}{(1)} & \multicolumn{1}{c}{(2)} & \multicolumn{1}{c}{(3)} \\
\hline
\multicolumn{2}{l}{\textit{observed \OmegaCIV{} evolution}} & --- &
2.4 \\
\\

enhanced & no delay &  1.5 &  0.7\\
enhanced & Chabrier  &  1.9 &  0.7\\
enhanced & Kroupa  &  2.3 &  0.8\\
enhanced & 0.5 Gyr delta  &  4.5 &  1.1 \\
enhanced & Cha + 0.5 Gyr  &  6.2 &  1.3\\
enhanced & Krp + 0.5 Gyr  &  7.8 &  1.5\\
enhanced & 0.6 Gyr delta  &  5.9 &  1.4\\
enhanced & Cha + 0.6 Gyr  &  8.4 &  1.8\\
enhanced & Krp + 0.6 Gyr  & 10.7 &  2.1\\
enhanced & 0.7 Gyr  delta &  7.9 &  2.1 \\
enhanced & Cha + 0.7 Gyr  & 11.8 &  2.6\\
enhanced & Krp + 0.7 Gyr  & 15.7 &  3.1 \\
\\

fiducial & no delay &  3.0 &  0.8\\
fiducial & Chabrier  &  3.9 &  0.9\\
fiducial & Kroupa  &  4.7 &  1.0\\
fiducial & 0.5 Gyr delta  & 10.8 &  2.3\\
fiducial & Cha + 0.5 Gyr  & 16.3 &  2.7\\
fiducial & Krp + 0.5 Gyr  & 21.7 &  3.2\\
fiducial & 0.6 Gyr delta  & 15.6 &  4.0\\
fiducial & Cha + 0.6 Gyr  & 24.7 &  4.8\\
fiducial & Krp + 0.6 Gyr  & 34.6 &  5.4\\
fiducial & 0.7 Gyr  delta & 25.4 &  8.2 \\
fiducial & Cha + 0.7 Gyr  & 42.7 &  9.5\\
fiducial & Krp + 0.7 Gyr  & 62.5 & 10.5 \\
\hline
\end{tabular}\medskip

(1) Figure \ref{fig_delayedenrich} shows the corresponding enrichment
history for each row in this table (except the $0.6$~Gyr
delays).\medskip

(2) The $f_{xCIV}$ value given is the one needed to match \OmegaCIV{}
at $z=4.7$. Compare to our estimates of $f_{xCIV} = 1.4 \times
10^{-9}$ for $Z = 1/50~Z_{\sun}$, or $5.0 \times 10^{-9}$ for $Z =
Z_{\sun}$.\medskip

(3) $\Delta\OmegaCIV{}/\OmegaCIV$ is independent of $f_{xCIV}$.

\end{table}

\begin{figure}
\includegraphics[width=3in]{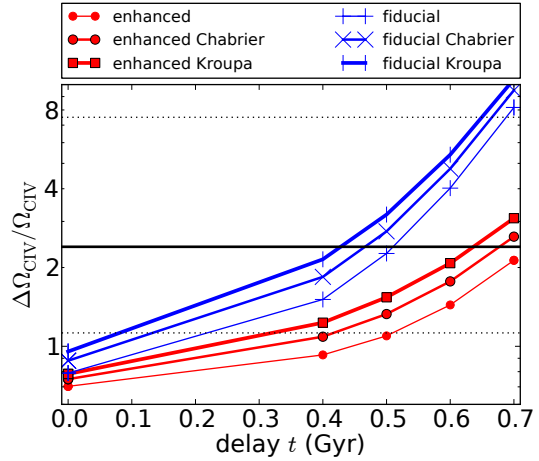}
\caption{\label{fig_deltas} Fractional increase in \ion{C}{iv} density
  as a function of the additional delay $t$. Note the log scale of the
  vertical axis. The bottom three curves show the fractional increase
  in \CIV{} density from $z=5.8$ to $4.7$ for the enhanced luminosity
  function history with no stellar--lifetime delay, the
  Chabrier--IMF--based delay, and the Kroupa--IMF--based delay (bottom
  to top). The top three curves show the corresponding increases with
  the fiducial LF history. The delay $t$ is \textit{in addition} to
  any stellar--lifetime delay. The solid horizontal line indicates the
  observed value of the increase ($2.4$), while the dotted lines show
  the ($1\sigma$) confidence limits on $\OmegaCIV(z=5.8)$. The curves
  with the additional stellar--lifetime delay (top two in each set)
  match the observed value at $t=0.4$ to $0.7$~Gyr, depending on the
  LF and IMF.}
\end{figure}

Figure \ref{fig_delayedenrich} shows enrichment histories with
$f_{xCIV}$ adjusted to fit $\OmegaCIV(z=4.7)$ to the observed
value. This allows us to assess the effect of the delay on the slope
by comparing the earlier evolution of the \OmegaCIV{} curve to the
observed value at $z=5.8$. If \ion{C}{iv} absorber production is too
slow, $\Delta \OmegaCIV/\OmegaCIV$ will be too small, and we will
over--predict the earlier measurement. Table \ref{tab_increase} gives
the $f_{xCIV}$ and $\Delta \OmegaCIV/\OmegaCIV$ values for each
combination of LF history and delay function. In Figure
\ref{fig_deltas}, we plot $\Delta \OmegaCIV/\OmegaCIV$ as a function
of the delay.

The stellar--lifetime delays alone are insufficient to boost the
\ion{C}{iv} growth to the observed rate, whereas assuming a $0.7$~Gyr
delay instead brings the enhanced LF history into agreement with the
observations (top panel of Figure \ref{fig_delayedenrich}). Since the
stellar--lifetime delay is inevitable, even if a longer delay is also
in effect, we next add the $0.7$~Gyr delay to the
stellar--lifetime--based delay (middle panel of Figure
\ref{fig_delayedenrich}). When the stellar--lifetime delays are added,
the fractional growth is increased by $25$--$50\%$ (for the Chabrier
and Kroupa delays, respectively). In the bottom panel of Figure
\ref{fig_delayedenrich} we add a shorter $0.5$~Gyr delay. Table
\ref{tab_increase} also includes an intermediate $0.6$~Gyr additional
delay. Figure \ref{fig_deltas} allows an easy comparison between the
predicted and observed values of $\Delta \OmegaCIV/\OmegaCIV$.

The stellar--lifetime delays have an effect on both the length of the
additional delay and the $f_{xCIV}$ value needed to match the
observations. For instance, with the enhanced LF history and a delta
function delay of $0.7$~Gyr, the fractional increase is $2.1$ and
$f_{xCIV}=7.9\times10^{-9}$. With the Kroupa--IMF delay included, a
shorter additional delay of $0.6$~Gyr produces the same fractional
increase, but then $f_{xCIV}$ must be increased by $35\%$. Therefore,
future models designed to study the relationships between
reionization, enrichment, and galaxy formation must include the finite
stellar lifetimes in order to draw precise quantitative conclusions.

While some of the $f_{xCIV}$ values in Table \ref{tab_increase} are
physically plausible, some may be unphysically high, indicating that
certain delay function and LF history combinations are not reasonable
candidates for explaining the observed increase in \ion{C}{iv}
density. For instance, the fiducial LF with a delta--function delay of
$0.7$~Gyr requires the rather high value of $f_{xCIV} = 2.5 \times
10^{-8}$. On the other hand, $f_{xCIV} = 7.9 \times 10^{-9}$ (for the
enhanced LF history with $0.7$~Gyr delay function delay) could
plausibly be explained as the result of a solar-metallicity $r_{\gamma
  Z}$ (ionizing photon to metal nucleon ratio) and $f_{\mathrm{esc}Z}
= 0.3$ (instead of $0.2$).

So far in this section, we have ignored the observational
uncertainties on \OmegaCIV. We are less concerned about the error at
$z=4.7$, since \OmegaCIV is approximately constant at $z \la
4.7$. Therefore in Figure \ref{fig_deltas}, we indicate the range of
the fractional increase corresponding to the $1\sigma$ confidence
interval on \OmegaCIV{} at $z=5.8$ (dotted lines). The current large
observational errors allow a wide range of values for the delay
(e.g. $1\sigma$ limits of $\sim 0.1$--$0.7$~Gyr for the fiducial LF),
but Figure \ref{fig_deltas} makes it clear that the current
constraints exclude zero delay at more than the $1\sigma$ level, and
that a combination of tighter constraints on $\OmegaCIV(z=5.8)$ and
the galaxy luminosity function at $z > 6$ has the potential to place
interesting constraints on the delay.

In Figure \ref{fig_deltas}, we have shown that, without any change in
the \ion{C}{iv} ionization correction, a $\sim 0.4$--$0.7$~Gyr delay
between the production of ionizing photons and \ion{C}{iv} absorption
features can explain the rapid increase in \ion{C}{iv} density between
$z=5.8$ and $4.7$. Simulations
\citep{2009MNRAS.396..729O,arXiv:1005.1451} suggest that the
triply--ionized fraction of carbon may increase by a factor of up to
$1.25$ in this interval, which would help to explain the observed
rise, but would leave a factor of $\ga 2.7$ increase ($\Delta \OmegaC
/ \OmegaC = 1.7$) in the total carbon content of the IGM. Figure
\ref{fig_deltas} tells us that this growth still requires a delay of
$\sim 0.6$~Gyr with the enhanced LF history, or $\sim 0.4$~Gyr for the
fiducial LF. In the next section we will explore possible physical
mechanisms for such delays.

\section{Discussion}\label{sec_discussion}

Since stellar lifetimes were too short to provide the required delay,
what mechanism could explain a longer $\sim 0.4$--$0.7$~Gyr timescale?
We show in this section that changing the stellar initial mass
function is not a viable explanation, but that galactic outflow
timescales correspond nicely to the required delay. We then make
testable predictions for two different outflow--driven delay
scenarios.

\subsection{Failed Explanations for the Delay: the IMF}\label{sec_delayimf}

One way to produce a longer mean delay in carbon production would be
to change the stellar IMF to increase the proportion of long--lived,
low--mass stars.%
\footnote{Though this is outside the stellar mass range we are
  concerned with, \citealt{arXiv:1009.5992} have found evidence for
  $\phi(m) \propto m^{-3}$ at $m \la 1~m_{\sun}$ in massive local
  early-type galaxies.} 
To produce the long delay times found above requires an IMF that is
radically bottom--heavy at high redshift. Even with a steep power--law
IMF of $\phi(m) \propto m^{-3.5}$, we find $\Delta \OmegaCIV /
\OmegaCIV < 1.1$. Therefore we also explored using a truncated
Salpeter IMF ($\phi(m) \propto m^{-2.35}$) with different maximum
masses, and found that a maximum stellar mass of $2.44~m_{\sun}$ (with
the fiducial LF) or $2.09~m_{\sun}$ (with the enhanced LF) is required
to produce a $\Delta \OmegaCIV / \OmegaCIV$ of $2.4$. These stellar
masses correspond to lifetimes of $0.4$ and $0.6$~Gyr, respectively,
in agreement with the delay times that we found above. However, these
IMFs would produce no ionizing radiation to accomplish reionization,
and would not match the constraints on the SEDs of high redshift
galaxies. Furthermore, a bottom--heavy IMF is in opposition to the
trend expected for stars forming from metal--poor gas (but see
\citealt*{2008ApJ...686..801O}, who propose that when the metallicity
exceeds a critical value of $\sim 10^{-5} Z_{\sun}$, dense clusters of
low-mass stars may form in the nuclei of second--generation
galaxies). Therefore changes to the IMF seem totally unable to explain
the rapid \OmegaCIV{} evolution.

\subsection{Explanations for the Delay: Galactic Outflows}\label{sec_delayoutflow}

In order to explain a long delay between star formation and
\ion{C}{iv} absorber production, we may posit that carbon ejected from
a galaxy must reach a characteristic distance before producing a
\ion{C}{iv} absorption feature visible in current data sets. In this
case, the relevant timescale for a distance $d$ and velocity $v$ would
be
\begin{equation}
t = 0.5 ~ \mathrm{Gyr} ~ \frac{d}{100~\mathrm{kpc}} \frac{200~\mathrm{km}~\mathrm{s}^{-1}}{v}.
\end{equation}
The characteristic distance from the source galaxy at which
\ion{C}{iv} is observed can be affected both by the ionization state
of the carbon as a function of distance, and the filling factor of
observable \ion{C}{iv} absorbers. If the volume filling factor of the
absorbers is too low, absorption systems will become too rare for
detection in the limited set of sight lines currently available.

While the exact relationship between galaxies and absorbers is still
highly uncertain (and contested), observations at $z = 2$--$3$ seem
consistent with this outflow scenario. \citet{2010ApJ...717..289S}
studied absorption associated with Lyman--break galaxies, both along
the line of sight to the galaxies and along the line of sight to
background galaxies at impact parameters from
$3$--$125~\mathrm{kpc}$. They found that the absorber equivalent width
in their composite spectra declines slowly with impact parameter
($W_0(\mathrm{CIV}) \propto b^{-0.2}$) up to $b \sim 80~\mathrm{kpc}$,
after which it quickly drops below their detection limit ($W_0 \sim
0.1$~\AA). The absorbers detected by \citet{2009MNRAS.395.1476R} at
$z=5.8$ range from $W_0 = 0.06$ to $\sim 0.7$~\AA, which would place
them at impact parameters from $b > 60$ to $b >
100~\mathrm{kpc}$. They fit the $W_0$ versus $b$ profile with a model
of a spherically-symmetric outflow with a covering fraction by
absorbing clouds of $f_\mathrm{cov}(r) \propto r^{-0.23}$. The slow
decline in covering fraction with distance indicates that the
absorbing clouds must be expanding as they move away from the source
galaxy (otherwise the covering fraction would go like $r^{-2}$). For a
constant expansion velocity and conserved cloud number $R_\mathrm{C}
\propto r^{1-\gamma/2}$ when $f_\mathrm{cov} \propto r^{-\gamma}$. The
inferred cloud radius therefore increases as $R_\mathrm{C} \propto
r^{0.9}$. The fact that the \CIV{} equivalent width declines more
slowly than other ions (for which $f_\mathrm{cov} \propto r^{-0.6}$)
may indicate that ionization effects are also serving to increase the
\CIV$/$C ratio with distance from the source galaxy. Therefore, these
observations suggest that both an increasing filling factor of
absorbers and an increasing triply--ionized fraction of carbon with
distance from the source galaxy could delay the appearance of \CIV{}
absorption in the IGM.

\citet{2010ApJ...717..289S} also use their direct line of sight
(impact parameter $b=0$) absorption profiles to constrain models of
the outflow velocity. They reproduce the observed profiles with an
accelerating velocity profile (higher velocities farther from the
source galaxy) reaching maximum velocities of $\sim
800~\mathrm{km}~\mathrm{s}^{-1}$. Given the $\sim 100~\mathrm{kpc}$
distance corresponding to absorbers with the equivalent width of the
\citet{2009MNRAS.395.1476R} sample, this gives a time scale of $\ga
0.1~\mathrm{Gyr}$. With the higher sensitivity of quasar spectra (but
using the absorber-absorber correlation function rather than direct
detection of the associated galaxies), \citet{2010ApJ...721..174M}
find a larger size for absorption regions of $\sim 150~\mathrm{kpc}$,
though they also suggest that the size may be constant in comoving
coordinates, which brings the physical size back down to
$85$--$100~\mathrm{kpc}$ at $z=6$--$5$.

\cite{2010ApJ...721..174M} and \citet{2009MNRAS.392.1539T}, using the
distortion between the line--of--sight (redshift space) and transverse
correlation functions of quasar spectrum absorbers constrain the
peculiar velocity of the average absorber to $\la
200~\mathrm{km}~{s}^{-1}$.  However, these analyses compared the mean
or central velocities of the absorbers. \citet{2010ApJ...717..289S}
showed that the absorption profiles are quite precisely centred on the
redshift of the source galaxy (at least at low impact parameters and
averaged over the angular size of the background galaxy), with the
\textit{width} of the profile extending to high velocities. Therefore
these results would constrain only the peculiar velocities of the host
galaxies, not the outflow velocities.

Both \citet{2006AJ....131...24S} and \citet{2007A&A...473..791F}
measured the velocity widths of \CIV{} absorbers, and found them to be
correlated with absorber column density. In the former sample,
absorbers with column densities comparable to the $z=5.8$ sample have
widths (at one-tenth maximum), of
$20$--$300~\mathrm{km}~\mathrm{s}^{-1}$, while in the later (a study
of DLA- and sub-DLA-associated absorption), all of the absorbers with
$\log N(\mathrm{CIV}) \ga 13.6$ have widths of
$>100~\mathrm{km}~\mathrm{s}^{-1}$, and most are
$200~\mathrm{km}~\mathrm{s}^{-1}$--$400~\mathrm{km}~\mathrm{s}^{-1}$.

Numerical hydrodynamic simulations also seem to suggest that both
ionization and filling--factor effects are at
work. \citet{2009MNRAS.396..729O} find that \ion{C}{iv} absorbers at
$z=6$ tend to have left their source galaxy $0.1$--$0.5$~Gyr earlier,
and lie within $10$--$50$~kpc of a galaxy in their simulation (giving
average speeds of $\sim 30$--$300~\mathrm{km}/\mathrm{s}$). They
suggest that the ionizing radiation from the source galaxy largely
determines the distance at which carbon is seen as
\ion{C}{iv}. \citet{arXiv:1005.1451} found absorbers at similar
distances ($\sim70$kpc). In their simulations, collisional ionization is
important to the \ion{C}{iv} ionization balance, and they suggest that
at least some \ion{C}{iv} absorbers are produced in shocks formed in
galactic outflows.

\citet{2009MNRAS.396..729O} also find that, as metals travel farther
from the source galaxy, they enrich less dense regions of the IGM and
produce weaker \ion{C}{iv} absorption systems. Therefore (neglecting
the ionization effects for the moment) absorbers near a galaxy will
have a higher column density, but a smaller filling factor (and will
therefore be detected more rarely), while absorbers farther from the
galaxy will have a lower column densities, but larger filling factors.

Whatever is determining the distance scale, a $\sim 0.4$--$0.7$~Gyr
delay could naturally be explained as the result of $\sim
200~\mathrm{km}/\mathrm{s}$ outflows carrying carbon to distances of
$\sim 80$--$140$~kpc before it is seen in \ion{C}{iv} absorption.

\subsection{Predictions}\label{sec_predictions}

We have shown that a delay between the production of ionizing photons
and of (currently-) detectable \ion{C}{iv} absorbers in the IGM can
solve the puzzles posed by the rapid rise in \OmegaCIV{} between $z
\approx 6$ and $5$, and have suggested that the delay could be due to
either (or to a combination) of two mechanisms associated with
galactic outflows. Each mechanism leads to distinct observable
predictions.

If the filling--factor evolution dominates the delay, it leads to an
interesting prediction: if the \ion{C}{iv} has indeed already been
ejected from galaxies, and the reason it is not yet seen at the
highest redshifts is that is has not yet spread to occupy a detectable
filling factor, then future observations, with significantly larger
effective lines--of--sight, should uncover much of this hitherto
hidden carbon in rare, high-column density \ion{C}{iv} absorbers (at
low impact parameters to unseen galaxies).%
\footnote{The lack of detection of weak \ion{C}{iv} absorbers in the
  \citet{2009ApJ...698.1010B} observations is evidence against the
  alternative hypothesis that the carbon is hidden in systems too weak
  to detect with most current spectra.}
This model would predict that, once the rare, high-column-density
end of the column density distribution has been probed, the total
integrated \OmegaCIV{} curve will be a scaled version of the dashed
curves in Figure \ref{fig_delayedenrich}, and the \OmegaCII{} curve
will be a scaled version of the \OmegaCIV{} curve. Together, these
observations would confirm that the amount of \ion{C}{iv} present in
the IGM tracks the cosmic star--formation history, but the rise in the
filling factor of \ion{C}{iv} systems is driven on a longer timescale,
determined by the finite (relatively low) speed of the
carbon--transporting winds or outflows. The evolution of the \CIV{}
column-density distribution and correlation function will provide
constraints on outflow and enrichment models.

On the other hand if ionization effects (either local or universal)
determine the delay, then ongoing and future measurements of the
density of \ion{C}{ii} should reveal that much of the missing carbon
is hidden in that ionization state. Similarly, observations of
\ion{Si}{ii}, \ion{Si}{iii}, and \ion{Si}{iv} absorbers should reveal
a changing ionization balance in that element.

These are clear and feasible ways to discriminate observationally
between ionization--driven evolution and
outflow--filling--factor--driven evolution. In either case, a careful
accounting of the total amount of carbon in the IGM (summing over
ionization states and the absorber column density distribution) will
reveal whether it tracks the cosmic star--formation history. Any
departure from a constant proportionality either indicates a problem
with measurements of the star--formation history, or a changing
efficiency of carbon production, both of which would be of
considerable interest.

\section{Conclusions}\label{sec_conclusions}

We have shown that a $\sim 0.4$--$0.7$~Gyr delay between the
production of ionizing photons and \ion{C}{iv} absorption features can
explain the rapid evolution of the \ion{C}{iv} density in the IGM
between $z=5.8$ and $4.7$. No change in the ratio of carbon to
ionizing photon production (the ionizing efficiency) or in the
universal ionization correction for \ion{C}{iv} is required. This
delay has a natural physical explanation, namely the need to transport
carbon a certain distance into the IGM before it is seen in
\ion{C}{iv} absorption. The distance scale would be determined by a
combination of the need to enrich a sufficient volume of the IGM to be
detectable in a limited number of quasar sight lines, and possible
ionization effects that optimise the \ion{C}{iv} fraction at a certain
distance from the source galaxy. An outflow of
$200~\mathrm{km}~\mathrm{s}^{-1}$ would carry material to a distance
of $80$--$140$~kpc on these timescales.

The shorter delay due to finite stellar lifetimes cannot provide the
full explanation for the rapid evolution (even with a steep stellar
initial mass function), but must be included in future models in order
to properly understand the relationships between galaxy formation,
reionization, and enrichment of the IGM.

Future measurements of metal abundance in the IGM will provide
important constraints on these relationships. Perhaps the most
important improvement will come from probing more quasar lines of
sight, both to reduce the uncertainty in the density of absorbers in
the currently--detected column--density range
\citep[$10^{13.5}$--$10^{14.5}$;][]{2009MNRAS.395.1476R}, and to
search for the rare, high--column--density absorbers that we expect to
exist close to galaxies. This search does not require extraordinarily
high resolution or signal--to--noise ratio
\citep{2009MNRAS.396..729O}, just the identification of more high--$z$
quasars and the taking of their near--IR absorption spectra (though
probing the weak end of the column--density distribution will require
higher quality spectra). The Multi Unit Spectroscopic Explorer
\citep[MUSE;][]{arXiv:astro-ph/0606329}, an integral field unit for
the VLT, will be ideal for directly exploring the relationship between
galaxies and IGM absorbers because it will be able to simultaneously
obtain spectroscopic redshifts for large numbers of galaxies (up to
$z\sim6.6$) near a quasar line of sight, which can then be compared
with the redshift distribution of absorbers seen in the quasar
spectrum.

Observations of multiple ionization stages of several elements, using
the sensitivity and wavelength coverage of new and future near--IR
spectrographs like X-shooter, will break the degeneracy of the factors
entering into $f_{xCIV}$, since some factors should be (approximately)
independent of element (such as $f_{\mathrm{esc}Z}$ and $r_{\gamma
  Z}$), while the fraction of carbon in the \ion{C}{iv} stage
($f_\mathrm{CIV}$) can be constrained from measurements of \ion{C}{ii}
absorption and of the ionization balance in other species
\citep{2009MNRAS.395.1476R}. Detailed physical models will be crucial
in interpreting these future observations. Such work is already
underway \citep{2009MNRAS.396..729O,arXiv:1005.1451}, but needs to be
improved to model both reionization and enrichment self--consistently.

\section{Acknowledgements}

We would like to thank Francesca Matteucci for discussion of stellar
chemical yields. ZH acknowledges financial support by the Pol\'anyi
Program of the Hungarian National Office for Research and Technology
(NKTH). RHK is supported by a Zwicky Fellowship at ETH Zurich. PM
would like to acknowledge NSF grant AST-0908910. We thank our
anonymous reviewer for useful and constrictive feedback.

This work made use of the excellent VizieR database of astronomical catalogues%
\footnote{\url{http://vizier.u-strasbg.fr/}}
\citep*{2000A&AS..143...23O}.

\bibliography{KHM2010}

\begin{thebibliography}{}

\bibitem[\protect\citeauthoryear{{Asplund}, {Grevesse} \& {Sauval}}{{Asplund}
  et~al.}{2005}]{2005ASPC..336...25A}
{Asplund} M.,  {Grevesse} N.,    {Sauval} A.~J.,  2005, in {T.~G.~Barnes III \&
  F.~N.~Bash} ed., Cosmic Abundances as Records of Stellar Evolution and
  Nucleosynthesis Vol.~336 of Astronomical Society of the Pacific Conference
  Series, {The Solar Chemical Composition}.
pp 25--+

\bibitem[\protect\citeauthoryear{{Bacon} et~al.,}{{Bacon}
  et~al.}{2006}]{arXiv:astro-ph/0606329}
{Bacon} R.,  et~al., 2006, The Messenger, 124, 5,
  \eprint{arXiv:astro-ph/0606329}

\bibitem[\protect\citeauthoryear{{Becker}, {Rauch} \& {Sargent}}{{Becker}
  et~al.}{2009}]{2009ApJ...698.1010B}
{Becker} G.~D.,  {Rauch} M.,    {Sargent} W.~L.~W.,  2009, \apj, 698, 1010,
  \eprint{arXiv:0812.2856}

\bibitem[\protect\citeauthoryear{{Bolton} \& {Haehnelt}}{{Bolton} \&
  {Haehnelt}}{2007}]{2007MNRAS.382..325B}
{Bolton} J.~S.,  {Haehnelt} M.~G.,  2007, \mnras, 382, 325,
  \eprint{arXiv:astro-ph/0703306}

\bibitem[\protect\citeauthoryear{{Bouwens}, {Illingworth}, {Franx} \&
  {Ford}}{{Bouwens} et~al.}{2007}]{2007ApJ...670..928B}
{Bouwens} R.~J.,  {Illingworth} G.~D.,  {Franx} M.,    {Ford} H.,  2007, \apj,
  670, 928, \eprint{arXiv:0707.2080}

\bibitem[\protect\citeauthoryear{{Bouwens}, {Illingworth}, {Franx} \&
  {Ford}}{{Bouwens} et~al.}{2008}]{2008ApJ...686..230B}
{Bouwens} R.~J.,  {Illingworth} G.~D.,  {Franx} M.,    {Ford} H.,  2008, \apj,
  686, 230, \eprint{arXiv:0803.0548}

\bibitem[\protect\citeauthoryear{{Bouwens}, {Illingworth}, {Oesch}, {Labbe},
  {Trenti}, {van Dokkum}, {Franx}, {Stiavelli}, {Carollo}, {Magee} \&
  {Gonzalez}}{{Bouwens} et~al.}{2010}]{arXiv:1006.4360}
{Bouwens} R.~J.,  {Illingworth} G.~D.,  {Oesch} P.~A.,  {Labbe} I.,  {Trenti}
  M.,  {van Dokkum} P.,  {Franx} M.,  {Stiavelli} M.,  {Carollo} C.~M.,
  {Magee} D.,    {Gonzalez} V.,  2010, ArXiv e-prints, \eprint{arXiv:1006.4360}

\bibitem[\protect\citeauthoryear{{Cen} \& {Chisari}}{{Cen} \&
  {Chisari}}{2010}]{arXiv:1005.1451}
{Cen} R.,  {Chisari} N.~E.,  2010, ArXiv e-prints, \eprint{arXiv:1005.1451}

\bibitem[\protect\citeauthoryear{{Chabrier}}{{Chabrier}}{2003}]{2003PASP..115.%
.763C}
{Chabrier} G.,  2003, \pasp, 115, 763, \eprint{arXiv:astro-ph/0304382}

\bibitem[\protect\citeauthoryear{{Chary}}{{Chary}}{2008}]{2008ApJ...680...32C}
{Chary} R.,  2008, \apj, 680, 32, \eprint{arXiv:0712.1498}

\bibitem[\protect\citeauthoryear{{Dav{\'e}} \& {Oppenheimer}}{{Dav{\'e}} \&
  {Oppenheimer}}{2007}]{2007MNRAS.374..427D}
{Dav{\'e}} R.,  {Oppenheimer} B.~D.,  2007, \mnras, 374, 427,
  \eprint{arXiv:astro-ph/0608268}

\bibitem[\protect\citeauthoryear{{Fox}, {Ledoux}, {Petitjean} \&
  {Srianand}}{{Fox} et~al.}{2007}]{2007A&A...473..791F}
{Fox} A.~J.,  {Ledoux} C.,  {Petitjean} P.,    {Srianand} R.,  2007, \aap, 473,
  791, \eprint{arXiv:0707.4065}

\bibitem[\protect\citeauthoryear{{Gavil{\'a}n}, {Buell} \&
  {Moll{\'a}}}{{Gavil{\'a}n} et~al.}{2005}]{2005A&A...432..861G}
{Gavil{\'a}n} M.,  {Buell} J.~F.,    {Moll{\'a}} M.,  2005, \aap, 432, 861,
  \eprint{arXiv:astro-ph/0411746}

\bibitem[\protect\citeauthoryear{{Hui} \& {Gnedin}}{{Hui} \&
  {Gnedin}}{1997}]{1997MNRAS.292...27H}
{Hui} L.,  {Gnedin} N.~Y.,  1997, \mnras, 292, 27,
  \eprint{arXiv:astro-ph/9612232}

\bibitem[\protect\citeauthoryear{{Kennicutt}
  Jr.}{{Kennicutt}}{1998}]{1998ARA&A..36..189K}
{Kennicutt} Jr. R.~C.,  1998, \araa, 36, 189, \eprint{arXiv:astro-ph/9807187}

\bibitem[\protect\citeauthoryear{{Komatsu} et~al.,}{{Komatsu}
  et~al.}{2010}]{arXiv:1001.4538}
{Komatsu} E.,  et~al., 2010, ArXiv e-prints, \eprint{arXiv:1001.4538}

\bibitem[\protect\citeauthoryear{{Kroupa}, {Tout} \& {Gilmore}}{{Kroupa}
  et~al.}{1993}]{1993MNRAS.262..545K}
{Kroupa} P.,  {Tout} C.~A.,    {Gilmore} G.,  1993, \mnras, 262, 545

\bibitem[\protect\citeauthoryear{{Martin}, {Scannapieco}, {Ellison}, {Hennawi},
  {Djorgovski} \& {Fournier}}{{Martin} et~al.}{2010}]{2010ApJ...721..174M}
{Martin} C.~L.,  {Scannapieco} E.,  {Ellison} S.~L.,  {Hennawi} J.~F.,
  {Djorgovski} S.~G.,    {Fournier} A.~P.,  2010, \apj, 721, 174,
  \eprint{arXiv:1007.2457}

\bibitem[\protect\citeauthoryear{{Mesinger}}{{Mesinger}}{2010}]{2010MNRAS.tmp.%
1017M}
{Mesinger} A.,  2010, \mnras, 407, 1328, \eprint{arXiv:0910.4161}

\bibitem[\protect\citeauthoryear{{Ochsenbein}, {Bauer} \&
  {Marcout}}{{Ochsenbein} et~al.}{2000}]{2000A&AS..143...23O}
{Ochsenbein} F.,  {Bauer} P.,    {Marcout} J.,  2000, \aaps, 143, 23,
  \eprint{arXiv:astro-ph/0002122}

\bibitem[\protect\citeauthoryear{{Oesch}, {Carollo}, {Stiavelli}, {Trenti},
  {Bergeron}, {Koekemoer}, {Lucas}, {Pavlovsky}, {Beckwith}, {Dahlen},
  {Ferguson}, {Gardner}, {Lilly}, {Mobasher} \& {Panagia}}{{Oesch}
  et~al.}{2009}]{2009ApJ...690.1350O}
{Oesch} P.~A.,  {Carollo} C.~M.,  {Stiavelli} M.,  {Trenti} M.,  {Bergeron}
  L.~E.,  {Koekemoer} A.~M.,  {Lucas} R.~A.,  {Pavlovsky} C.~M.,  {Beckwith}
  S.~V.~W.,  {Dahlen} T.,  {Ferguson} H.~C.,  {Gardner} J.~P.,  {Lilly} S.~J.,
  {Mobasher} B.,    {Panagia} N.,  2009, \apj, 690, 1350,
  \eprint{arXiv:0804.4874}

\bibitem[\protect\citeauthoryear{{Omukai}, {Schneider} \& {Haiman}}{{Omukai}
  et~al.}{2008}]{2008ApJ...686..801O}
{Omukai} K.,  {Schneider} R.,    {Haiman} Z.,  2008, \apj, 686, 801,
  \eprint{arXiv:0804.3141}

\bibitem[\protect\citeauthoryear{{Oppenheimer} \& {Dav{\'e}}}{{Oppenheimer} \&
  {Dav{\'e}}}{2006}]{2006MNRAS.373.1265O}
{Oppenheimer} B.~D.,  {Dav{\'e}} R.,  2006, \mnras, 373, 1265,
  \eprint{arXiv:astro-ph/0605651}

\bibitem[\protect\citeauthoryear{{Oppenheimer} \& {Dav{\'e}}}{{Oppenheimer} \&
  {Dav{\'e}}}{2008}]{2008MNRAS.387..577O}
{Oppenheimer} B.~D.,  {Dav{\'e}} R.,  2008, \mnras, 387, 577,
  \eprint{arXiv:0712.1827}

\bibitem[\protect\citeauthoryear{{Oppenheimer}, {Dav{\'e}} \&
  {Finlator}}{{Oppenheimer} et~al.}{2009}]{2009MNRAS.396..729O}
{Oppenheimer} B.~D.,  {Dav{\'e}} R.,    {Finlator} K.,  2009, \mnras, 396, 729,
  \eprint{arXiv:0901.0286}

\bibitem[\protect\citeauthoryear{{Pawlik}, {Schaye} \& {van
  Scherpenzeel}}{{Pawlik} et~al.}{2009a}]{arXiv:0912.3034}
{Pawlik} A.~H.,  {Schaye} J.,    {van Scherpenzeel} E.,  2009a, ArXiv e-prints,
  \eprint{arXiv:0912.3034}

\bibitem[\protect\citeauthoryear{{Pawlik}, {Schaye} \& {van
  Scherpenzeel}}{{Pawlik} et~al.}{2009b}]{2009MNRAS.394.1812P}
{Pawlik} A.~H.,  {Schaye} J.,    {van Scherpenzeel} E.,  2009b, \mnras, 394,
  1812, \eprint{arXiv:0807.3963}

\bibitem[\protect\citeauthoryear{{Pettini}, {Madau}, {Bolte}, {Prochaska},
  {Ellison} \& {Fan}}{{Pettini} et~al.}{2003}]{2003ApJ...594..695P}
{Pettini} M.,  {Madau} P.,  {Bolte} M.,  {Prochaska} J.~X.,  {Ellison} S.~L.,
   {Fan} X.,  2003, \apj, 594, 695, \eprint{arXiv:astro-ph/0305413}

\bibitem[\protect\citeauthoryear{{Rollinde}, {Vangioni}, {Maurin}, {Olive},
  {Daigne}, {Silk} \& {Vincent}}{{Rollinde} et~al.}{2009}]{2009MNRAS.398.1782R}
{Rollinde} E.,  {Vangioni} E.,  {Maurin} D.,  {Olive} K.~A.,  {Daigne} F.,
  {Silk} J.,    {Vincent} F.~H.,  2009, \mnras, 398, 1782,
  \eprint{arXiv:0806.2663}

\bibitem[\protect\citeauthoryear{{Romano}, {Chiappini}, {Matteucci} \&
  {Tosi}}{{Romano} et~al.}{2005}]{2005A&A...430..491R}
{Romano} D.,  {Chiappini} C.,  {Matteucci} F.,    {Tosi} M.,  2005, \aap, 430,
  491

\bibitem[\protect\citeauthoryear{{Romano}, {Karakas}, {Tosi} \&
  {Matteucci}}{{Romano} et~al.}{2010}]{arXiv:1006.5863}
{Romano} D.,  {Karakas} A.~I.,  {Tosi} M.,    {Matteucci} F.,  2010, ArXiv
  e-prints, \eprint{arXiv:1006.5863}

\bibitem[\protect\citeauthoryear{{Ryan-Weber}, {Pettini}, {Madau} \&
  {Zych}}{{Ryan-Weber} et~al.}{2009}]{2009MNRAS.395.1476R}
{Ryan-Weber} E.~V.,  {Pettini} M.,  {Madau} P.,    {Zych} B.~J.,  2009, \mnras,
  395, 1476, \eprint{arXiv:0902.1991}

\bibitem[\protect\citeauthoryear{{Salvaterra}, {Ferrara} \&
  {Dayal}}{{Salvaterra} et~al.}{2010}]{arXiv:1003.3873}
{Salvaterra} R.,  {Ferrara} A.,    {Dayal} P.,  2010, ArXiv e-prints,
  \eprint{arXiv:1003.3873}

\bibitem[\protect\citeauthoryear{{Schaerer}}{{Schaerer}}{2002}]{2002A&A...382.%
..28S}
{Schaerer} D.,  2002, \aap, 382, 28, \eprint{arXiv:astro-ph/0110697}

\bibitem[\protect\citeauthoryear{{Songaila}}{{Songaila}}{2001}]{2001ApJ...561L%
.153S}
{Songaila} A.,  2001, \apjl, 561, L153, \eprint{arXiv:astro-ph/0110123}

\bibitem[\protect\citeauthoryear{{Songaila}}{{Songaila}}{2006}]{2006AJ....131.%
..24S}
{Songaila} A.,  2006, \aj, 131, 24, \eprint{arXiv:astro-ph/0509821}

\bibitem[\protect\citeauthoryear{{Steidel}, {Erb}, {Shapley}, {Pettini},
  {Reddy}, {Bogosavljevi{\'c}}, {Rudie} \& {Rakic}}{{Steidel}
  et~al.}{2010}]{2010ApJ...717..289S}
{Steidel} C.~C.,  {Erb} D.~K.,  {Shapley} A.~E.,  {Pettini} M.,  {Reddy} N.,
  {Bogosavljevi{\'c}} M.,  {Rudie} G.~C.,    {Rakic} O.,  2010, \apj, 717, 289,
  \eprint{arXiv:1003.0679}

\bibitem[\protect\citeauthoryear{{Tilvi}, {Rhoads}, {Hibon}, {Malhotra},
  {Wang}, {Veilleux}, {Swaters}, {Probst}, {Krug}, {Finkelstein} \&
  {Dickinson}}{{Tilvi} et~al.}{2010}]{arXiv:1006.3071}
{Tilvi} V.,  {Rhoads} J.~E.,  {Hibon} P.,  {Malhotra} S.,  {Wang} J.,
  {Veilleux} S.,  {Swaters} R.,  {Probst} R.,  {Krug} H.,  {Finkelstein} S.~L.,
     {Dickinson} M.,  2010, \apj, 721, 1853, \eprint{arXiv:1006.3071}

\bibitem[\protect\citeauthoryear{{Trac} \& {Cen}}{{Trac} \&
  {Cen}}{2007}]{2007ApJ...671....1T}
{Trac} H.,  {Cen} R.,  2007, \apj, 671, 1, \eprint{arXiv:astro-ph/0612406}

\bibitem[\protect\citeauthoryear{{Tyson}}{{Tyson}}{1988}]{1988ApJ...329L..57T}
{Tyson} N.~D.,  1988, \apjl, 329, L57

\bibitem[\protect\citeauthoryear{{Tytler}, {Gleed}, {Melis}, {Chapman},
  {Kirkman}, {Lubin}, {Paschos}, {Jena} \& {Crotts}}{{Tytler}
  et~al.}{2009}]{2009MNRAS.392.1539T}
{Tytler} D.,  {Gleed} M.,  {Melis} C.,  {Chapman} A.,  {Kirkman} D.,  {Lubin}
  D.,  {Paschos} P.,  {Jena} T.,    {Crotts} A.~P.~S.,  2009, \mnras, 392, 1539

\bibitem[\protect\citeauthoryear{{van Dokkum} \& {Conroy}}{{van Dokkum} \&
  {Conroy}}{2010}]{arXiv:1009.5992}
{van Dokkum} P.,  {Conroy} C.,  2010, ArXiv e-prints, \eprint{arXiv:1009.5992}

\bibitem[\protect\citeauthoryear{{Ventura} \& {Marigo}}{{Ventura} \&
  {Marigo}}{2010}]{arXiv:1007.2533}
{Ventura} P.,  {Marigo} P.,  2010, \mnras, pp 1247--+, \eprint{arXiv:1007.2533}

\bibitem[\protect\citeauthoryear{{Woosley} \& {Weaver}}{{Woosley} \&
  {Weaver}}{1995}]{1995ApJS..101..181W}
{Woosley} S.~E.,  {Weaver} T.~A.,  1995, \apjs, 101, 181

\end{thebibliography}

\end{document}